\catcode`@=11
\magnification \magstep1
\tolerance=1500\frenchspacing
\input psfig.sty
%
%

\font\titfnt=cmbx10 scaled \magstep1
\font\pagfnt=cmsl10
\font\parfnt=cmbx10 scaled \magstep1
\font\sprfnt=cmsl10 scaled \magstep1
\font\teofnt=cmssbx10
\font\tenenufnt=cmsl10
\font\nineenufnt=cmsl9

\font\ninerm=cmr9
\font\eightrm=cmr8
\font\sixrm=cmr6
 
\font\ninei=cmmi9
\font\eighti=cmmi8
\font\sixi=cmmi6
\skewchar\ninei='177 \skewchar\eighti='177
\skewchar\sixi='177
 
\font\ninesy=cmsy9
\font\eightsy=cmsy8
\font\sixsy=cmsy6
\skewchar\ninesy='60 \skewchar\eightsy='60
\skewchar\sixsy='60

\font\ninebf=cmbx9

\font\sixbf=cmbx6
\font\fivebf=cmbx5
 
\font\ninett=cmtt9
\font\eighttt=cmtt8
 
\font\tenmat=cmss10

\font\sevenmat=cmss7 
 
\font\fivemat=cmss5
\newfam\matfam

\hyphenchar\tentt=-1 
\hyphenchar\ninett=-1
\hyphenchar\eighttt=-1
 
\font\ninesl=cmsl9

\font\nineit=cmti9

\font\tengrm=gsmn1000

\font\sevengrm=gsmn0700

\font\fivegrm=gsmn0500
\newfam\grmfam

  \def\grm{\fam\grmfam\tengrm} \textfont\grmfam=\tengrm
     \scriptfont\grmfam=\sevengrm \scriptscriptfont\grmfam=\fivegrm
 
\newskip\ttglue
%
%
\input wasyfont.tex
%
%
\newdimen\pagewidth \newdimen\pageheight
\newdimen\ruleht
\hsize=31.6pc  \vsize=44.5pc  \maxdepth=2.2pt
\parindent=19pt
\parindent=19pt
\hfuzz=2pt
\pagewidth=\hsize \pageheight=\vsize \ruleht=.5pt
\abovedisplayskip=6pt plus 3pt minus 1pt
\belowdisplayskip=6pt plus 3pt minus 1pt
\abovedisplayshortskip=0pt plus 3pt
\belowdisplayshortskip=4pt plus 3pt
\def\blank{\vskip 12pt}
\def\blankii{\blank\blank}

\def\blankq{\vskip 3pt}
%
%
\def\tenpoint{\def\rm{\fam0\tenrm}%
  \def\enufnt{\tenenufnt}%
  \textfont0=\tenrm \scriptfont0=\sevenrm
\scriptscriptfont0=\fiverm
  \textfont1=\teni \scriptfont1=\seveni
\scriptscriptfont1=\fivei
  \textfont2=\tensy \scriptfont2=\sevensy
\scriptscriptfont2=\fivesy
  \textfont3=\tenex \scriptfont3=\tenex
\scriptscriptfont3=\tenex
  \def\it{\fam\itfam\tenit}%
  \textfont\itfam=\tenit
  \def\sl{\fam\slfam\tensl}%
  \textfont\slfam=\tensl
  \def\bf{\fam\bffam\tenbf}%
  \textfont\bffam=\tenbf \scriptfont\bffam=\sevenbf
   \scriptscriptfont\bffam=\fivebf
  \def\mat{\fam\matfam\tenmat} \textfont\matfam=\tenmat
     \scriptfont\matfam=\sevenmat \scriptscriptfont\matfam=\fivemat
  \def\tt{\fam\ttfam\tentt}%
  \textfont\ttfam=\tentt
  \tt \ttglue=.5em plus.25em minus.15em
  \normalbaselineskip=12pt
  \let\sc=\eightrm
  \let\big=\tenbig
  \setbox\strutbox=\hbox{\vrule height8.5pt depth3.5pt width\z@}%
  \normalbaselines\rm}
 
\def\ninepoint{\def\rm{\fam0\ninerm}%
  \def\enufnt{\nineenufnt}%
  \textfont0=\ninerm \scriptfont0=\sixrm
\scriptscriptfont0=\fiverm
  \textfont1=\ninei \scriptfont1=\sixi
\scriptscriptfont1=\fivei
  \textfont2=\ninesy \scriptfont2=\sixsy
\scriptscriptfont2=\fivesy
  \textfont3=\tenex \scriptfont3=\tenex
\scriptscriptfont3=\tenex
  \def\it{\fam\itfam\nineit}%
  \textfont\itfam=\nineit
  \def\sl{\fam\slfam\ninesl}%
  \textfont\slfam=\ninesl
  \def\bf{\fam\bffam\ninebf}%
  \textfont\bffam=\ninebf \scriptfont\bffam=\sixbf
   \scriptscriptfont\bffam=\fivebf
  \def\tt{\fam\ttfam\ninett}%
  \textfont\ttfam=\ninett
  \tt \ttglue=.5em plus.25em minus.15em
  \normalbaselineskip=11pt
  \let\sc=\sevenrm
  \let\big=\ninebig
  \setbox\strutbox=\hbox{\vrule height8pt depth3pt width\z@}%
  \normalbaselines\rm}
 
\def\tenbig#1{{\hbox{$\left#1\vbox to8.5pt{}\right.\n@space$}}}
\def\ninebig#1{{\hbox{$\textfont0=\tenrm\textfont2=\tensy
  \left#1\vbox to7.25pt{}\right.\n@space$}}}
\def\eightbig#1{{\hbox{$\textfont0=\ninerm\textfont2=\ninesy
  \left#1\vbox to6.5pt{}\right.\n@space$}}}
%
%
\def\corsivo{\enufnt}

\newcount\cont@note
\global\cont@note=0
\def\footnote{\global\advance\cont@note by 1
      \edef\@sf{\spacefactor\the\spacefactor}$^{\the\cont@note}$\@sf
      \insert\footins\bgroup\ninepoint
      \interlinepenalty100 \let\par=\endgraf
        \leftskip=\z@skip \rightskip=\z@skip
        \splittopskip=10pt plus 1pt minus 1pt \floatingpenalty=20000
        \smallskip\item{$^{\the\cont@note}$}\bgroup\strut\aftergroup\@foot\let\next}
\skip\footins=12pt plus 2pt minus 4pt 
\dimen\footins=30pc 
%
%
\newcount\numbibliogr@fi@
\global\numbibliogr@fi@=1
\newwrite\fileref
\immediate\openout\fileref=\jobname.ref
\immediate\write\fileref{\parindent 30pt}
\def\cita#1#2{\def\us@gett@{\the\numbibliogr@fi@}
\expandafter\xdef\csname:bib_#1\endcsname{\us@gett@}
           \immediate\write\fileref
           {\par\noexpand\item{{[\the\numbibliogr@fi@]\enspace}}}\ignorespaces
           \immediate\write\fileref{{#2}}\ignorespaces
           \global\advance\numbibliogr@fi@ by 1\ignorespaces}
\def\bibref#1{\seindefinito{:bib_#1}
          \immediate\write16{ !!! \string\bibref{#1} non definita !!!}
\expandafter\xdef\csname:bib_#1\endcsname{??}\fi
      {$^{[\csname:bib_#1\endcsname]}$}}
\def\dbiref#1{\seindefinito{:bib_#1}
          \immediate\write16{ !!! \string\bibref{#1} non definita !!!}
\expandafter\xdef\csname:bib_#1\endcsname{??}\fi
      {[\csname:bib_#1\endcsname]}}

\def\references{\immediate\closeout\fileref
                \par\goodbreak
                \blankii
                \centerline{\parfnt References}
                \nobreak\blank\nobreak
                \input \jobname.ref}

%
%
\def\title#1{\null\blankii\noindent{\titfnt\uppercase{#1}}\blank}

\def\author#1{\leftskip 1.8cm\smallskip\noindent{#1}\smallskip\leftskip 0pt}
\def\abstract#1{\par\blankii\noindent
          {\ninepoint
            {\bf Abstract. }{\rm #1}}
          \par}

%
%
%
\def\seindefinito#1{\expandafter\ifx\csname#1\endcsname\relax}
%
\newdimen\@mpiezz@
\@mpiezz@=\hsize
\newbox\boxp@r@gr@fo
\def\section#1#2{
           \goodbreak\vskip 18pt plus 6pt\noindent\ignorespaces
              {\setbox\boxp@r@gr@fo=\hbox{\parfnt\noindent\ignorespaces
              {\secref{#1}.\quad}}\ignorespaces
             \advance\@mpiezz@ by -\wd\boxp@r@gr@fo
             \box\boxp@r@gr@fo\vtop{\hsize=\@mpiezz@\noindent\parfnt #2}}
           \par\nobreak\vskip 9pt plus 3pt\nobreak
           \noindent\ignorespaces}
\def\secref#1{\seindefinito{:sec_#1}
          \immediate\write16{ !!! \string\secref{#1} non definita !!!}
 \expandafter\xdef\csname:sec_#1\endcsname{??}\fi
      \csname:sec_#1\endcsname
      }
%
\newcount\num@ppendice
\global\num@ppendice=64
\def\appendix#1#2{
         \goodbreak\vskip 18pt plus 6pt\noindent\ignorespaces
              {\setbox\boxp@r@gr@fo=\hbox{\parfnt\noindent\ignorespaces
              {\appref{#1}.\quad}}\ignorespaces
             \advance\@mpiezz@ by -\wd\boxp@r@gr@fo
             \box\boxp@r@gr@fo\vtop{\hsize=\@mpiezz@\noindent\parfnt #2}}
           \par\nobreak\vskip 9pt plus 3pt\nobreak
           \noindent\ignorespaces}
\def\appref#1{\seindefinito{:app_#1}
          \immediate\write16{ !!! \string\appref{#1} non definita !!!}
          \expandafter\xdef\csname:app_#1\endcsname{??}\fi
      {\csname:app_#1\endcsname}}
%
\def\subsection#1#2{
           \goodbreak\vskip 9pt plus 2pt\noindent\ignorespaces
           {\setbox\boxp@r@gr@fo=\hbox{\sprfnt\noindent\ignorespaces
              {\sbsref{#1}\quad}}\ignorespaces
             \advance\@mpiezz@ by -\wd\boxp@r@gr@fo
         \box\boxp@r@gr@fo\vtop{\hsize=\@mpiezz@\noindent\sprfnt#2}}
           \par\nobreak\vskip 3pt plus 1pt\nobreak
           \noindent\ignorespaces}
\def\sbsref#1{\seindefinito{:sbs_#1}
          \immediate\write16{ !!! \string\sbsref{#1} non definita !!!}
 \expandafter\xdef\csname:sbs_#1\endcsname{??}\fi
      \csname:sbs_#1\endcsname
      }
%

\def\formdef#1{\relax}
\def\formula#1{\leqno{(\csname:frm_#1\endcsname)}}
\def\frmref#1{\seindefinito{:frm_#1}
          \immediate\write16{ !!! \string\frmref{#1} non definita !!!}
 \expandafter\xdef\csname:frm_#1\endcsname{??}\fi
      (\csname:frm_#1\endcsname)}
%
\def\endclaim{\endgroup
            \par\if F\sp@zi@tur@{\blankq}\gdef\sp@zi@tur@{T}\fi}
\def\theorem#1{\par\if T\sp@zi@tur@{\gdef\sp@zi@tur@{F}}\else{\blankq}\fi
            \noindent{\teofnt Theorem \thrref{#1}:\quad}\begingroup\enufnt
            \ignorespaces}
\def\theoremnn{
            \par\if T\sp@zi@tur@{\gdef\sp@zi@tur@{F}}\else{\blankq}\fi
            \noindent{\teofnt Theorem:\quad}\begingroup\enufnt
            \ignorespaces}
\def\theoremtx#1#2{\par\if T\sp@zi@tur@{\gdef\sp@zi@tur@{F}}\else{\blankq}\fi
            \noindent{\teofnt Theorem \thrref{#1}@:
                {\enufnt #2}.\quad}\begingroup\enufnt
            \ignorespaces}
\def\thrref#1{\seindefinito{:thr_#1}
          \immediate\write16{ !!! \string\thrref{#1} non definita !!!}
 \expandafter\xdef\csname:thr_#1\endcsname{??}\fi
      \csname:thr_#1\endcsname}
%
\def\proposition#1{\par\if T\sp@zi@tur@{\gdef\sp@zi@tur@{F}}\else{\blankq}\fi
            \noindent{\teofnt Proposition \proref{#1}:\quad}\begingroup\enufnt
            \ignorespaces}
\def\propositiontx#1#2{\par\if T\sp@zi@tur@{\gdef\sp@zi@tur@{F}}\else{\blankq}\fi
            \noindent{\teofnt Proposition \proref{#1}:
               {\enufnt #2}.\quad}\begingroup\enufnt
            \ignorespaces}
\def\proref#1{\seindefinito{:pro_#1}
          \immediate\write16{ !!! \string\proref{#1} non definita !!!}
 \expandafter\xdef\csname:pro_#1\endcsname{??}\fi
      \csname:pro_#1\endcsname}
%
\def\corollary#1{\par\if T\sp@zi@tur@{\gdef\sp@zi@tur@{F}}\else{\blankq}\fi
            \noindent{\teofnt Corollary \corref{#1}:\quad}\begingroup\enufnt
            \ignorespaces}
\def\corref#1{\seindefinito{:cor_#1}
          \immediate\write16{ !!! \string\corref{#1} non definita !!!}
 \expandafter\xdef\csname:cor_#1\endcsname{??}\fi
      \csname:cor_#1\endcsname}
%
\def\lemma#1{\par\if T\sp@zi@tur@{\gdef\sp@zi@tur@{F}}\else{\blankq}\fi
            \noindent{\teofnt Lemma \lemref{#1}:\quad}\begingroup\enufnt
            \ignorespaces}
\def\lemref#1{\seindefinito{:lem_#1}
          \immediate\write16{ !!! \string\lemref{#1} non definita !!!}
 \expandafter\xdef\csname:lem_#1\endcsname{??}\fi
      \csname:lem_#1\endcsname}
%
\def\definition#1{\par\if T\sp@zi@tur@{\gdef\sp@zi@tur@{F}}\else{\blankq}\fi
            \noindent{\teofnt Definition \defref{#1}:\quad}\begingroup\enufnt
            \ignorespaces}
\def\definitiontx#1#2{\par\if T\sp@zi@tur@{\gdef\sp@zi@tur@{F}}\else{\blankq}\fi
            \noindent{\teofnt Definition \defref{#1}:
               {\enufnt #2}.\quad}\begingroup\enufnt
            \ignorespaces}
\def\defref#1{\seindefinito{:def_#1}
          \immediate\write16{ !!! \string\defref{#1} non definita !!!}
 \expandafter\xdef\csname:def_#1\endcsname{??}\fi
      \csname:def_#1\endcsname}
%
\def\proof{\par\if T\sp@zi@tur@{\gdef\sp@zi@tur@{F}}\else{\blankq}\fi
    \noindent{\teofnt Proof.\quad}\begingroup\ignorespaces}
\def\prooftx#1{\par\if T\sp@zi@tur@{\gdef\sp@zi@tur@{F}}\else{\blankq}\fi
    \noindent{\teofnt Proof #1.\quad}\begingroup\ignorespaces}

%
\newbox\boxfigur@
\newbox\comfigur@
\newdimen\@mpfigur@
\newdimen\m@rfigur@
\m@rfigur@=2 pc
\@mpfigur@=\hsize
\advance\@mpfigur@ by -2\m@rfigur@
\def\figure#1#2#3{
      \setbox\boxfigur@\vbox{\centerline{#2}}
      \topinsert
         {\vbox{
               \vskip 1pt
               \box\boxfigur@
               \vskip 4pt}}
      \setbox\comfigur@\vtop{\hsize=\@mpfigur@\parindent 0pt
         {\ninepoint
         {\teofnt Figure \csname:fig_#1\endcsname.}\enspace{#3}}
      }
      \centerline{\box\comfigur@}
      \endinsert
      \write16{Figure {\csname:fig_#1\endcsname}.}
      }
\def\figcont#1#2{
      \setbox\boxfigur@\vbox{\centerline{#2}}
      \topinsert
         {\vbox{
               \vskip 1pt
               \box\boxfigur@
               \vskip 4pt}}
 \setbox\comfigur@\vtop{\hsize=\@mpfigur@\parindent 0pt
         {\ninepoint
         {\teofnt Figure \figref{#1}.}\enspace{(continued).}}
      }
      \centerline{\box\comfigur@}
      \endinsert
      \write16{Figura (cont) {\csname:fig_#1\endcsname}.}
}
\def\figref#1{\seindefinito{:fig_#1}
          \immediate\write16{ !!! \string\figref{#1} non definita !!!}
 \expandafter\xdef\csname:fig_#1\endcsname{??}\fi
      \csname:fig_#1\endcsname}
\def\citazione{\par\begingroup\corsivo
     \everypar{\parshape 1 \m@rfigur@ \@mpfigur@} 
     \noindent\llap{``\enspace}\ignorespaces}
\def\finecitazione{\rlap{\enspace ''}\par\endgroup}
%
\def\remarks{\par\if
T\sp@zi@tur@{\gdef\sp@zi@tur@{F}}\else{\blankq}\fi
    \noindent{\teofnt Remarks.\quad}\ignorespaces}
\def\remark{\par\if
T\sp@zi@tur@{\gdef\sp@zi@tur@{F}}\else{\blankq}\fi
    \noindent{\teofnt Remark.\quad}\ignorespaces}
\newbox\boxt@vol@
\newbox\comt@vol@
\newdimen\@mpt@vol@
\newdimen\m@rt@vol@
\m@rt@vol@=2 pc
\@mpt@vol@=\hsize
\advance\@mpt@vol@ by -2\m@rt@vol@
\def\table#1#2#3{
      \setbox\boxt@vol@\vbox{#3}
      \topinsert
      \setbox\comt@vol@\vtop{\hsize=\@mpt@vol@\parindent 0pt
         {\ninepoint
         {\teofnt Table \tabref{#1}.}\enspace{#2}}
      }
      \centerline{\box\comt@vol@}
      \vskip 5pt%
      \centerline{\box\boxt@vol@}%
      \endinsert
      \write16{Table {\csname:tav_#1\endcsname}.}
}
\def\tabcont#1#2{
      \setbox\boxt@vol@\vbox{#2}
      \topinsert
      \setbox\comt@vol@\vtop{\hsize=\@mpt@vol@\parindent 0pt
         {\ninepoint
         {\teofnt Table \tabref{#1}.}\enspace{(continued)}}
      }
      \centerline{\box\comt@vol@}
      \vskip 5pt%
      \centerline{\box\boxt@vol@}%
      \endinsert
      \write16{Table {\csname:tav_#1\endcsname}.}
}
\def\tabref#1{\seindefinito{:tav_#1}
          \immediate\write16{ !!! \string\tabref{#1} non definita !!!}
          \expandafter\xdef\csname:tav_#1\endcsname{??}\fi
      \csname:tav_#1\endcsname}
%
%

\def\noblank{\gdef\sp@zi@tur@{T}}
\def\incolonna#1{\displ@y \tabskip=\centering
   \halign to \displaywidth{\hfil$\@lign\displaystyle{{}##}$\tabskip=0pt
       &$\@lign\displaystyle{{}##}$\hfil\tabskip=\centering
       &\llap{$\@lign\displaystyle{{}##}$}\tabskip=0pt\crcr
       #1\crcr}}
%
%
\nopagenumbers
\def\testos{\null}
\def\testod{\null}
\headline={\if T\tpage{\gdef\tpage{F}{\hfil}}
 \else{\ifodd\pageno\rightheadline\else\leftheadline\fi}
           \fi}
 
\gdef\tpage{T}
\def\rightheadline{\hfil{\pagfnt\testod}\hfil{\pagfnt\folio}}
\def\leftheadline{{\pagfnt\folio}\hfil{\pagfnt\testos}\hfil}
\voffset=2\baselineskip
\everypar={\gdef\sp@zi@tur@{F}}
\catcode`@=12

\def\pmb#1{\setbox0=\hbox{#1}\ignorespaces
    \hbox{\kern-.02em\copy0\kern-\wd0\ignorespaces
    \kern.05em\copy0\kern-\wd0\ignorespaces
    \kern-.02em\raise.02em\box0 }}

\def\rho{\varrho}
\def\theta{\vartheta}
\def\phi{\varphi}
\def\epsilon{\varepsilon}

\def\frac#1#2{{{#1}\over{#2}}}

\tenpoint\rm


\def\altriga{\phantom{\vrule width 0pt height 11pt depth 5pt}}
\expandafter\edef\csname:bib_Euler-1748\endcsname{1}
\expandafter\edef\csname:bib_Euler-1752\endcsname{2}
\expandafter\edef\csname:bib_Haretu-1885\endcsname{3}
\expandafter\edef\csname:bib_Giorgilli-1998\endcsname{4}
\expandafter\edef\csname:bib_Gio-Loc-2009\endcsname{5}
\expandafter\edef\csname:bib_Keplero-1609\endcsname{6}
\expandafter\edef\csname:bib_Keplero-cons\endcsname{7}
\expandafter\edef\csname:bib_Keplero-tavole\endcsname{8}
\expandafter\edef\csname:bib_Kolmogorov-1954\endcsname{9}
\expandafter\edef\csname:bib_Kremer-1980\endcsname{10}
\expandafter\edef\csname:bib_Kremer-1981\endcsname{11}
\expandafter\edef\csname:bib_Lagrange-1762\endcsname{12}
\expandafter\edef\csname:bib_Lagrange-1774\endcsname{13}
\expandafter\edef\csname:bib_Lagrange-1781\endcsname{14}
\expandafter\edef\csname:bib_Lagrange-1782\endcsname{15}
\expandafter\edef\csname:bib_Laplace-1773\endcsname{16}
\expandafter\edef\csname:bib_Laplace-1785\endcsname{17}
\expandafter\edef\csname:bib_Laskar-1989\endcsname{18}
\expandafter\edef\csname:bib_Laskar-1994\endcsname{19}
\expandafter\edef\csname:bib_Laskar-2006\endcsname{20}
\expandafter\edef\csname:bib_Laskar-2009\endcsname{21}
\expandafter\edef\csname:bib_Schoner-1544\endcsname{22}
\expandafter\edef\csname:bib_Wilson-1985\endcsname{23}
\expandafter\edef\csname:sec_1\endcsname{1}
\expandafter\edef\csname:tav_oss.saturno\endcsname{1}
\expandafter\edef\csname:tav_oss.giove\endcsname{2}
\expandafter\edef\csname:sec_kepsec.1\endcsname{2}
\expandafter\edef\csname:fig_osserv.2\endcsname{1}
\expandafter\edef\csname:tav_oss.marte\endcsname{3}
\expandafter\edef\csname:tav_oss.venere\endcsname{4}
\expandafter\edef\csname:tav_oss.mercurio\endcsname{5}
\expandafter\edef\csname:fig_osserv.3\endcsname{2}
\expandafter\edef\csname:sec_kepsec.2\endcsname{3}
\expandafter\edef\csname:fig_kepsec.1\endcsname{3}
\expandafter\edef\csname:sec_kepsec.4\endcsname{4}

\cita{Euler-1748}{L.\ Euler: {\it Recherches sur la question des
in\'egalit\'es du mouvement de Saturne et de Jupiter}, Piece qui a
remport\'e le prix de l'Acad\'emie Royale des Sciences, 1--123
(1748). Reprinted in: Opera Omnia, Ser.\ 2, Vol.\ 25, 45--157.}

\cita{Euler-1752}{L.\ Euler: {\it Recherches sur les
inegalit\'es du mouvement de Jupiter et Saturne}, Recueil des
pieces qui ont remport\'e les prix de l'Acad\'emie Royale des Sciences
{\bf 7}, 1769, pp. 2-84. Reprinted in: Opera Omnia, Ser.\ 2, Vol.\
26.}

\cita{Haretu-1885}{S.C.\ Haretu: {\it Sur l'invariabilit\'e des
grands axes des orbites plan\'etaires}, Ann.\ Obs.\ Paris, M\'emoires,
{\bf 18}, 1--39 (1885).}

\cita{Giorgilli-1998}{A.\ Giorgilli: {\it Small denominators and
exponential stability: from Poincar\'e to the present time},
Rend. Sem. Mat. Fis. Milano, {\bf LXVIII}, 19--57 (1998).}

\cita{Gio-Loc-2009}{A.\ Giorgilli, U.\ Locatelli: {\it Sulla
stabilit\`a del problema planetario dei tre corpi}, Istituto Lombardo
-- Accademia di Scienze e Lettere, Rendiconti -- Classe di Scienze,
{\bf 143} (2009).}

\cita{Keplero-1609}{{\it Astronomia Nova}, seu Physica C{\oe}lestis
tradita commentariis de motibus Stell{\ae} Martis ex observationibus
G.V.\ Tychonis Brahe; Jussu \& sumptibus Rudolphi II, Romanorum
Imperatoris \&c.  Plurium annorum pertinaci studio elaborata
Prag{\ae}, A.S.C.M.S.\ Mathematico Johanne Keplero, cum ejusdem
C.M. privilegio speciali, Anno {\ae}r{\ae} Dionysian{\ae} MDCIX.
Reprinted in: {\it Johannis Kepleri astronomi opera omnia}, edidit
Dr.\ Ch.\ Frisch, Frankfurti A.M.\ et Erlang{\ae} Heyder \& Zimmer,
MDCCCLX, Vol.\ III.}

\cita{Keplero-cons}{J.\ Kepler: {\it Consideratio observationum
Regiomontani et Waltheri}, in: {\it Johannis Kepleri astronomi opera
omnia}, edidit Dr.\ Ch.\ Frisch, Frankfurti A.M.\ et Erlang{\ae} Heyder
\& Zimmer, MDCCCLX, Vol.\ VI, pp.\ 725--774.}

\cita{Keplero-tavole}{J.\ Kepler: {\it In tabulas Rudolphi
pr{\ae}fatio}, in: {\it Johannis Kepleri astronomi opera omnia},
edidit Dr.\ Ch.\ Frisch, Frankfurti A.M.\ et Erlang{\ae} Heyder \&
Zimmer, MDCCCLX, Vol.\ VI, pp.\ 666--674.}

\cita{Kolmogorov-1954}{A.N. Kolmogorov: {\it Preservation of
conditionally periodic movements with small change in the Hamilton
function}, Dokl. Akad. Nauk SSSR, {\bf 98}, 527 (1954).  English
translation in: Los Alamos Scientific Laboratory translation
LA-TR-71-67; reprinted in: Lecture Notes in Physics {\bf 93}.}

\cita{Kremer-1980}{R.L.\ Kremer: {\it Bernhard Walther's astronomical
observations}, Journal for the history of astronomy {\bf xi}, 174--191
(1980).}

\cita{Kremer-1981}{R.L.\ Kremer: {\it The use of Bernhard Walther's
astronomical observations: theory and observation in early modern
astronomy}, Journal for the history of astronomy {\bf xii}, 124--132
(1981).}

\cita{Lagrange-1762}{J.L. Lagrange: {\it Solution de diff\'erents
probl\`emes de calcul int\'egral}, Miscellanea Taurinensia,
Tomo III (1762--1765).  Reprinted in: {\it Oeuvres de
Lagrange}, Gauthier--Villars, Paris (1870), tome I, p.471--668.}

\cita{Lagrange-1774}{J.L.\ Lagrange: {\it Recherche sur les \'equations
s\'eculaires des mouvements des noeuds et des inclinaisons des orbites
des plan\`etes}, M\'emoires de l'Acad\'emie
Royale des Sciences de Paris (1774).   Reprinted in: {\it Oeuvres de
Lagrange}, Gauthier--Villars, Paris (1870), tome VI, p.635--709.}

\cita{Lagrange-1781}{J.L.\ Lagrange: {\it Th\'eorie des variations
s\'eculaires des \'el\'ements des plan\`etes. Premi\`ere partie
contenant les principes et les formules g\'en\'erales pour
d\'eterminer ces variations}, Nouveaux m\'emoires de l'Acad\'emie des
Sciences et Belles--Lettres de Berlin (1781).  Reprinted in: {\it
Oeuvres de Lagrange}, Gauthier--Villars, Paris (1870), tome V,
p.125--207.}

\cita{Lagrange-1782}{J.L.\ Lagrange: {\it Th\'eorie des variations
s\'eculaires des \'el\'ements des plan\`etes. Se\-con\-de partie
contenant la d\'etermination de ces variations pour chacune des
plan\`etes pricipales}, Nouveaux m\'emoires de l'Acad\'emie des
Sciences et Belles--Lettres de Berlin (1782).  Reprinted in: {\it
Oeuvres de Lagrange}, Gauthier--Villars, Paris (1870), tome V,
p.211--489.}

\cita{Laplace-1773}{P--S.\ de Laplace: {\it M\'emoire sur le principe
de la gravitation universelle et sur les in\'egalit\'es s\'eculaires
des plan\`etes qui en dependent}, M\'emoires de l'Acad\'emie Royale
des Sciences de Paris (1773).  Reprinted in: {Oeuvres compl\`etes de
Laplace}, Gauthier--Villars, Paris (1891), tome VIII, p.201--275.}

\cita{Laplace-1785}{P--S.\ de Laplace: {\it Th\'eorie de Jupiter et
Saturne}, M\'emoires de l'Acad\'emie Royale des Sciences de Paris,
ann\'ee 1785, (1788).  Reprinted in: {Oeuvres compl\`etes de Laplace},
{\bf XI}, p. 95.}

\cita{Laskar-1989}{J. Laskar: {\it A numerical experiment on the
chaotic behaviour of the solar system}, Nature, {\bf 338}, 237--238
(1989).}

\cita{Laskar-1994}{J. Laskar: {\it Large scale chaos in the solar
system}, Astron. Astroph. {\bf 287} (1994).}

\cita{Laskar-2006}{J.\ Laskar: {\it Lagrange et la Stabilit\'e du
Syt\`eme Solaire}, in: G.\ Sacchi Landriani e A.\ Giorgilli (eds):
{\it Sfogliando la M\'echanique Analitique}, LED edizioni, Milano
(2008).}

\cita{Laskar-2009}{Laskar, J., Gastineau, M.: {\it Existence of
collisional trajectories of Mercury, Mars and Venus with the Earth},
Nature {\bf 459}, 817--819 (2009).}

\cita{Schoner-1544}{J.\ Sch\"oner (ed.): {\it Scripta clarissimi mathematici
M.\ Johannis Regiomontani}, Nuremberg (1544).  Republished in
facsimile in: J.\ Regiomontanus: {\it Opera collectanea}, F.\
Schmeidler ed., 567--752, Osnabr\"uck (1972).  Walter's observations
were also printed in: W.\ Snel: {\it Coeli et siderum in eo errantium
Hassiac{\ae}}, Leiden (1618) and in: L.\ Barettus [A.\ Curtz]: {\it
Historia coelestis}, Augsburg (1666).}

\cita{Wilson-1985}{C.\ Wilson: {\it The great inequality of Jupiter and
Saturn: from Kepler to Laplace},  Archive for History of Exact Sciences
{\bf 33}, 15--290, (1985).}

\noindent
To appear in: \hfill\break
{\sl Rendiconti dell'Istituto Lombardo Accademia di Scienze e Lettere,
Classe di Scienze.}

\title{A Kepler's note on secular inequalities}

\author{\it ANTONIO GIORGILLI
\hfill\break Dipartimento di Matematica, Universit\`a degli Studi di Milano,
\hfill\break via Saldini 50, 20133\ ---\ Milano, Italy.}

\abstract{I discuss the problem of secular inequalities in Kepler by
giving account of a manuscript note that has not been published until
1860.  In his note Kepler points out the need for a model, clearly
inspired by the method of epicycles, that describes the secular
inequalities as periodic ones.  I bring attention to this point, that
seems to have been underestimated, since the references to Kepler's
work usually report only that he observed a decreasing mean motion for
Saturn and an increasing one for Jupiter.}

\section{1}{Introduction}
The discovery by Kepler of the elliptic shape of the planetary orbits
is often considered as a discontinuity with the traditional models of
Classical Astronomy, based on geometrical tools such as circles,
epicycles and equants.  The enthusiastic announcement of the discovery
in chapter~LVI of {\corsivo Astronomia Nova}~\dbiref{Keplero-1609},
where Kepler says that he was {\corsivo ``quasi e somno expergefactus,
et novam lucem intuitus''} (like suddenly awakened from sleep, and
seeing a new light) seems to confirm that this was his feeling, too.
But, as often happens, reality turned out to be complex enough to
escape our theories: the orbits of the planets are not exactly
elliptic.  This is well known today, but it seems that the
circumstancy that long term deviations from the elliptic motion have
been investigated in great detail by Kepler himself, with an explicit
conjecture that these deviations should be periodic, is not so known,
even among astronomers.

Kepler's opinion is clearly stated at the beginning of the preface to
the {\corsivo Tabul{\ae} Rudolphin{\ae}}~\dbiref{Keplero-tavole},
where he says:\footnote{``And the observations made in our epoch,
especially by Brahe, will prove the certainty of our calculations.
However, concerning the future we can not expect so much.  The
validity may be questioned by ancient observations, that I'm well
aware of, by the knowledge of the mean motions, that have not yet been
fully explored, and by the concurrence of physical actions.  The
observations of Regiomontanus and Walther do indeed show that we
should definitely think about secular equations, as I will explain 
in a specially devoted booklet.  Which and how many equations we
need, however, makind will be unable to decide before many centuries
of observations have been passed.''}

\citazione
{\corsivo Et de certitudine quidem calculi testabuntur observationes
pr{\ae}sentium temporum, imprimis Brahean{\ae}; de futuris vero
temporibus plura pr{\ae}\-su\-me\-re non possumus, quam vel observationes
veterum, quibus usus sum, vel ipsa motuum mediorum conditio, nondum
penitus explorata, concursusque causarum physicarum pr{\ae}stare
possunt, cum observationes Regiomontani et Waltheri testentur, omnino
de {\ae}quationibus secularibus esse cogitandum, ut singulari libello
reddam demonstratum suo tempore; qu{\ae} tamen {\ae}quationes quales et
quant{\ae} sint, ante plurimum s{\ae}culorum decursum observationesque
eorum, a gente humana definiri nequaquam possunt.}
\finecitazione
\blankq

\noindent
The booklet promised by Kepler was never published by him, and
probably this caused the details about his work to be forgotten.  A
preliminary manuscript was found after his death, but it was not
published until 1860, when it was included in the complete edition of
Kepler's works~\dbiref{Keplero-cons}.  The present memoir intends to
give a short report on Kepler's note.\footnote{I should mention that
my attention on Kepler's note was prompted by a conference given by
J.~Laskar at the Istituto Lombardo Accademia di Scienze e
Lettere~\dbiref{Laskar-2006}.  The present note can be considered as
complementary to Laskar's memoir.}  A short account on further
developments after Kepler is also included.

\table{oss.saturno}{Observed positions of Saturn compared with
the calculated ones.  The first column reports the date of the
observation; the second one the longitude of the observed planet as
reconstructed by Kepler on the basis of the observational data by
Regiomontanus and Walther; the third one the longitude predicted by
Kepler; the fourth one the difference.  The word ``idem'' in the first
column means the Kepler gives two possible outcomes for his
determinations of the longitude of the planet.}{
\def\num#1{\obeyspaces{\oldstyle #1}}
\vbox{\tabskip=0pt\offinterlineskip
\parindent 0pt\tabskip 0pt\halign{
\vrule$\altriga\ $\hfil{#}\ \vrule
&$\ $\hfil$\displaystyle{#}$\hfil{\ }
&{#}\hfil\ \vrule
&$\ $\hfil$\displaystyle{#}$\hfil{\ }
&$\ ${#}\hfil\ \vrule
&$\ $\hfil$\displaystyle{#}$\hfil\ \vrule
\cr
\noalign{\vskip 12pt}
\noalign{\hrule}
Date\hfil  & {\rm Observed}  && {\rm Calculated} && {\rm Difference} \cr
\noalign{\hrule}
2.12.1461  & {\num{29}}^{\circ}\,{\num{38}}'                           & {\capricornus} 
           & {\num{29}}^{\circ}\,{\num{14}}'{\scriptstyle\frac{1}{2}}  & {\capricornus}
           & -\num{23}'{\scriptstyle\frac{1}{2}}
\cr
idem       & {\num{29}}^{\circ}\,\num{50}'                             & {\capricornus} 
           & {\num{29}}^{\circ}\,{\num{14}}'{\scriptstyle\frac{1}{2}}  & {\capricornus}
           & -\num{35}'{\scriptstyle\frac{1}{2}}
\cr
17.09.1475 & {\num{ 5}}^{\circ}\,{\num{38}}'{\scriptstyle\frac{1}{2}}  & {\leo}
           & {\num{ 4}}^{\circ}\,{\num{42}}'                           & {\leo}
           & -\num{56}'{\scriptstyle\frac{1}{2}}
\cr
idem       & {\num{ 5}}^{\circ}\,{\num{54}}'                           & {\leo}
           & {\num{ 4}}^{\circ}\,{\num{42}}'                           & {\leo}
           & -\num{1}^{\circ}\,\num{12}'
\cr
25.03.1476 & {\num{ 0}}^{\circ}\,{\num{49}}'                           & {\leo}
           & {\num{ 0}}^{\circ}\,{\num{6}}'                            & {\leo}
           & -\num{43}'
\cr
13.10.1476 & {\num{21}}^{\circ}\,{\num{32}}'                           & {\leo}
           & {\num{20}}^{\circ}\,{\num{46}}'{\scriptstyle\frac{1}{2}}  & {\leo}
           & -\num{45}'{\scriptstyle\frac{1}{2}}
\cr
9.10.1477  & {\num{ 2}}^{\circ}\,{\num{13}}'                           & {\virgo}
           & {\num{ 1}}^{\circ}\,{\num{23}}'                           & {\virgo}
           & -\num{50}'
\cr
idem       & {\num{ 2}}^{\circ}\,\num{28}'                             & {\virgo}
           & {\num{ 1}}^{\circ}\,{\num{23}}'                           & {\virgo}
           & -{\num{1}}^{\circ}\,{\num{5}}'
\cr
19.04.1478 & {\num{28}}^{\circ}\,{\num{10}}'                           & {\leo}
           & \num{27}^{\circ}\,\num{18}'                               & {\leo}
           & -\num{52}'
\cr
idem       & {\num{28}}^{\circ}\,{\num{36}}'                           & {\leo}
           & {\num{27}}^{\circ}\,\num{18}'                             & {\leo}
           & -\num{1}^{\circ}\,\num{18}'
\cr
24.09.1478 & {\num{12}}^{\circ}\,{\num{15}}'                           & {\virgo} 
           & {\num{11}}^{\circ}\,{\num{32}}'                           & {\virgo}
           & -\num{43}'
\cr
30.10.1479 & {\num{27}}^{\circ}\,{\num{28}}'{\scriptstyle\frac{1}{2}}  & {\virgo}
           & {\num{26}}^{\circ}\,{\num{50}}'                           & {\virgo}
           & -\num{38}'{\scriptstyle\frac{1}{2}}
\cr
22.10.1481 & {\num{21}}^{\circ}\,{\num{13}}'                           & {\libra}
           & {\num{20}}^{\circ}\,{\num{45}}'{\scriptstyle\frac{1}{2}}  & {\libra}
           & -\num{27}'{\scriptstyle\frac{1}{2}}
\cr
idem       & {\num{21}}^{\circ}\,{\num{25}}'                           & {\libra}
           & {\num{20}}^{\circ}\,{\num{45}}'{\scriptstyle\frac{1}{2}}  & {\libra}
           & -\num{39}'{\scriptstyle\frac{1}{2}}
\cr
12.01.1482 & {\num{25}}^{\circ}\,{\num{11}}'                           & {\libra}
           & {\num{24}}^{\circ}\,{\num{31}}'                           & {\libra}
           & -\num{40}'
\cr
21.11.1484 & {\num{23}}^{\circ}\,{\num{30}}'                           & {\scorpio}
           & {\num{22}}^{\circ}\,{\num{50}}'                           & {\scorpio}
           & -\num{40}'
\cr
15.10.1503 & {\num{17}}^{\circ}\,{\num{46}}'{\scriptstyle\frac{1}{2}}  & {\cancer}
           & {\num{17}}^{\circ}\,{\num{ 0}}'{\scriptstyle\frac{1}{2}}  & {\cancer}
           & -\num{46}'
\cr
idem       & {\num{17}}^{\circ}\,{\num{58}}'                           & {\cancer}
           & {\num{17}}^{\circ}\,{\num{0}}'{\scriptstyle\frac{1}{2}}   & {\cancer}
           & -\num{58}'{\scriptstyle\frac{1}{2}}
\cr
11.12.1503 & {\num{15}}^{\circ}\,{\num{30}}'                           & {\cancer}
           & {\num{14}}^{\circ}\,{\num{51}}'{\scriptstyle\frac{1}{2}}  & {\cancer}
           & -\num{38}'{\scriptstyle\frac{1}{2}}
\cr
idem       & {\num{15}}^{\circ}\,{\num{34}}'{\scriptstyle\frac{1}{2}}  & {\cancer}
           & {\num{14}}^{\circ}\,{\num{51}}'{\scriptstyle\frac{1}{2}}  & {\cancer}
           & -\num{43}'
\cr
8.02.1504  & {\num{11}}^{\circ}\,{\num{11}}'                           & {\cancer}
           & {\num{10}}^{\circ}\,{\num{29}}'                           & {\cancer}
           & -\num{42}'
\cr
idem       & {\num{11}}^{\circ}\,{\num{15}}'{\scriptstyle\frac{1}{2}}  & {\cancer}
           & {\num{10}}^{\circ}\,{\num{29}}'                           & {\cancer}
           & -\num{46}'{\scriptstyle\frac{1}{2}}
\cr
8.03.1504  & {\num{10}}^{\circ}\,{\num{51}}'                           & {\cancer}
           & {\num{10}}^{\circ}\,{\num{5}}'                            & {\cancer}
           & -\num{46}'
\cr
29.03.1504 & {\num{11}}^{\circ}\,{\num{34}}'                           & {\cancer}
           & {\num{10}}^{\circ}\,{\num{46}}'                           & {\cancer}
           & -\num{48}'
\cr
24.05.1504 & {\num{16}}^{\circ}\,{\num{19}}'                           & {\cancer}
           & {\num{15}}^{\circ}\,{\num{37}}'{\scriptstyle\frac{1}{2}}  & {\cancer}
           & -\num{41}'{\scriptstyle\frac{1}{2}}
\cr
24.02.1514 & {\num{25}}^{\circ}\,{\num{13}}'{\scriptstyle\frac{1}{3}}  & {\scorpio}
           & {\num{24}}^{\circ}\,{\num{56}}'{\scriptstyle\frac{2}{3}}  & {\scorpio}
           & -\num{16}'{\scriptstyle\frac{1}{3}}
\cr
\noalign{\hrule}
}}}

\table{oss.giove}{Observed positions of Jupiter compared with
the calculated ones.}{
\def\num#1{{\oldstyle #1}}
\vbox{\tabskip=0pt\offinterlineskip
\parindent 0pt\tabskip 0pt\halign{
\vrule$\altriga\ $\hfil{#}\ \vrule
&$\ $\hfil$\displaystyle{#}$\hfil{\ }
&{#}\hfil\ \vrule
&$\ $\hfil$\displaystyle{#}$\hfil{\ }
&$\ ${#}\hfil\ \vrule
&$\ $\hfil$\displaystyle{#}$\hfil\ \vrule
\cr
\noalign{\vskip 12pt}
\noalign{\hrule}
Date\hfil  & {\rm Observed}  && {\rm Calculated} && {\rm Difference} \cr
\noalign{\hrule}
20.03.1462 & {\num{2}}^{\circ}\,{\num{15}}'                           & {\capricornus}
           & {\num{2}}^{\circ}\,{\num{21}}'{\scriptstyle\frac{3}{4}}  & {\capricornus}
           & +{\num{6}}'{\scriptstyle\frac{3}{4}}
\cr				
26.04.1468 & {\num{29}}^{\circ}\,{\num{24}}'{\scriptstyle\frac{2}{3}} & {\gemini}
           & {\num{29}}^{\circ}\,{\num{24}}'{\scriptstyle\frac{2}{3}} & {\gemini}
           & {\num{0}}'
\cr				
29.04.1468 & {\num{29}}^{\circ}\,{\num{56}}'{\scriptstyle\frac{1}{2}} & {\gemini}
           & {\num{29}}^{\circ}\,{\num{56}}'{\scriptstyle\frac{1}{2}} & {\gemini}
           & {\num{0}}'
\cr				
15.03.1471 & {\num{29}}^{\circ}\,{\num{33}}'                          & {\virgo}
           & {\num{29}}^{\circ}\,{\num{16}}'                          & {\virgo}
           & +\num{17}'
\cr				
21.02.1478 & {\num{24}}^{\circ}\,{\num{33}}'                          & {\aries}
           & {\num{25}}^{\circ}\,{\num{0}}'                           & {\aries}
           & +\num{27}'
\cr				
22.08.1478 & {\num{0}}^{\circ}\,{\num{25}}'{\scriptstyle\frac{1}{2}}  & {\gemini}
           & {\num{0}}^{\circ}\,{\num{39}}'{\scriptstyle\frac{1}{2}}  & {\gemini}
           & +\num{14}'
\cr				
30.09.1478 & {\num{0}}^{\circ}\,{\num{24}}'                           & {\gemini}
           & {\num{0}}^{\circ}\,{\num{42}}'{\scriptstyle\frac{1}{2}}  & {\gemini}
           & +\num{18}'{\scriptstyle\frac{1}{2}}
\cr				
21.11.1484 & {\num{23}}^{\circ}\,{\num{33}}'                          & {\gemini}
           & {\num{23}}^{\circ}\,{\num{48}}'{\scriptstyle\frac{1}{2}}  & {\gemini}
           & +\num{15}'
\cr				
10.01.1488 & {\num{23}}^{\circ}\,{\num{51}}'{\scriptstyle\frac{1}{2}} & {\aquarius} 
           & {\num{24}}^{\circ}\,{\num{2}}'                           & {\aquarius}
           & +\num{10}'{\scriptstyle\frac{1}{2}}
\cr				
15.02.1497 & {\num{12}}^{\circ}\,{\num{38}}'{\scriptstyle\frac{3}{4}} & {\sagittarius}
           & {\num{12}}^{\circ}\,{\num{38}}'{\scriptstyle\frac{3}{4}} & {\sagittarius}
           & {\num{0}}'
\cr				
8.09.1503  & {\num{11}}^{\circ}\,{\num{37}}'                          & {\cancer}
           & {\num{11}}^{\circ}\,{\num{52}}'                          & {\cancer}
           & +\num{18}'
\cr				
19.01.1504 & {\num{6}}^{\circ}\,{\num{31}}'{\scriptstyle\frac{2}{3}}  & {\cancer}
           & {\num{6}}^{\circ}\,{\num{50}}'                           & {\cancer}
           & +\num{18}'{\scriptstyle\frac{1}{3}}
\cr				
9.02.1504  & {\num{4}}^{\circ}\,{\num{57}}'{\scriptstyle\frac{1}{2}}  & {\cancer}
           & {\num{5}}^{\circ}\,{\num{22}}'{\scriptstyle\frac{1}{3}}  & {\cancer}
           & +\num{24}'{\scriptstyle\frac{5}{6}}
\cr				
29.04.1504 & {\num{11}}^{\circ}\,{\num{37}}'                          & {\cancer}
           & {\num{11}}^{\circ}\,{\num{52}}'                          & {\cancer}
           & +\num{15}'
\cr				
24.05.1504 & {\num{16}}^{\circ}\,{\num{11}}'{\scriptstyle\frac{1}{2}} & {\cancer}
           & {\num{16}}^{\circ}\,{\num{35}}'                           & {\cancer}
           & +\num{23}'{\scriptstyle\frac{1}{2}}
\cr				
\noalign{\hrule}
}}}

\section{kepsec.1}{Data from ancient observations}
As the title page of the {\corsivo Tabul{\ae} Rudolphin{\ae}} says,
the idea of compiling new astronomical tables was conceived by Tycho
Brahe (1546--1601) since 1564 (he was 17 years old), and the actual
work was started after 1572.  Kepler began his collaboration with
Tycho Brahe in 1600, and after Brahe's death he undertook the job of
continuing the compilation.  He completed the tables by 1623, but only
on 1627 he could publish them.  The observational data on which
Kepler's work was based were largely due to his mentor Tycho Brahe,
who had succeeded in measuring the positions of celestial bodies with
a precision which he claimed to be $1'$ (actually about $2'$ in most
cases), a great achievement for his epoch.

As it appears from the sentence quoted in the Introduction, Kepler was
not just satisfied by the excellent agreement between the
computational predictions based on his tables and Tycho Brahe's data.
He had indeed undertaken a big work in order to compare his
calculations with ancient observations available to him.  This is the
origin of his handwritten note.

Most of the ancient data available to Kepler were due to Johannes
M\"uller der K\"onisberg (1436--1476), also named Regiomontanus, and
Bernard Walther (1430--1504).  They had collected a long series of
observations between 1461 and 1504 which were considered as the most
complete and precise at that time.  Some data are due to Ptolemy
(c.$\,$90--c.$\,$168), who also reports previous observations due to
Chaldean astronomers (made around 229~BC) and to Dionysius (made
around 270~BC).  One observation is due to Copernicus.

Let me give a few examples of visual observations by Regiomontanus and
Walther.  This will illustrate how Kepler had to work in order to use
them.  The first example refers to Jupiter.\footnote{``In 1478, august
22, tree hours past midnight, Jupiter and the two eyes of Taurus were
on the same line, and Jupiter was toward west, the distance from the
west eye of Taurus being half the distance between the two eyes; so it
appeared to the eye.''}

\citazione{Anno 1478, 22 Aug.~h.~3 post medium noctis fuerant in una
linea $\jupiter$ et duo oculi $\taurus$, et erat $\jupiter$
occidentalior, distans per medietatem distanti{\ae}, qua duo oculi
distant, ab oculo occidentaliori; sic visui apparuit.
}\finecitazione

\noindent
It should be remarked that this is an example of a quite precise
observation, since the position of Jupiter is given with reference to
two well identified fixed stars that can be found in a good catalog.
In other cases only the relative position of two or more moving
objects is given, typically conjunctions among two or more planets
and/or the Moon.\footnote{``In 1478, September 24, $40$~minutes before
sunrise I saw Saturn and the Moon approximately in conjunction; Moon's
position was scanty on the north direction, and it appeared that
between her circle and Saturn one could insert the width of a palm.''}

\citazione{1478, 24 Sept, $40'$ ante ortum solis vidi lunam circa
$\saturn$, quasi coniunctos; distabat Luna modicum ad septentrionem,
ita ut inter circunferentiam eius et $\saturn$ videretur mediare
spatium unius palm{\ae}.
}\finecitazione

\noindent
In this case Kepler determines the longitude of the Moon, which allows
him to evaluate the longitude of Saturn.  Sometimes the indications of
Regiomontanus are definitely more challenging.\footnote{``$(\ldots)\>$the star
Jupiter was seen between two stars of Virgo.  The brighter one is
close to the center of the left wing of Virgo, the other one, less
bright, is close to her eye, towards Leo$\>(\ldots)$''}

\citazione{$(\ldots)\>$stella {\jupiter} videbatur inter duas Virginis, quarum
lucidior est circa medietatem al{\ae} sinistr{\ae} Virginis, alia
obscurior circa oculum eius versus Leonem$\>(\ldots)$
}\finecitazione

\noindent
Kepler's comment is: {\corsivo ``Magna cum perplexitate diu
conflictatus sum, q{\ae}nam essent h{\ae}c du{\ae} stell{\ae}''}
(With great perplexity and for a long I racked my brain trying to
figure out which these two stars are).  Then he makes an hypothesis
that seems reasonable, and proceeds with the calculation.

After 1488 Walther began to observe with the help or zodiacal
armillary spheres, so he could give longitudes and latitudes in
degrees and minutes of arc.  He was also used to add notes concerning
the reliability of his observations.\footnote{The report of Walther's
observations has been published by Johann Sch\"oner in
1544 (see~\dbiref{Schoner-1544}).  A rather detailed account of the data
collected by Walther in 1503--1504 can be found
in~\dbiref{Kremer-1980} and~\dbiref{Kremer-1981}.}

Kepler actually made a major effort in reconstructing the reported
observations so as to find the longitudes and often also the latitudes
of the planets, which allowed him to make a comparison with the
tables.  He also devotes some pages to a discussion of the reliability
of the measured data, possibly affected by a bad setting of the
armillary spheres, or by incorrect time indications.  Sometimes he
points out the possible existence of typographical errors.  I will not
discuss this aspect here.

A synopsis of Kepler's results is reported in
tables~\tabref{oss.saturno}--\tabref{oss.mercurio}.  I collected only
data for the longitudes of the planets.\footnote{In Kepler's
notations, the circle of the ecliptic is divided into 12 sectors of
$30^{\circ}$ each, identified by the constellations, and the longitude
is given relative to the constellation.  E.g., the notation
$17^{\circ}\,13'\>\libra$ means the longitude $17^{\circ}\,13'$
measured from the west point of the constellation of Libra.  Following
Kepler, in the text and in the tables I use the common astronomical
symbols for the constellations.  For the reader's convenience I
include here a list: $\aries$ (Aries), $\taurus$ (Taurus), $\gemini$
(Gemini), $\cancer$ (Cancer), $\leo$ (Leo), $\virgo$ (Virgo), $\libra$
(Libra), $\scorpio$ (Scorpio), $\sagittarius$ (Sagittarius),
$\capricornus$ (Capricornus), $\aquarius$ (Aquarius), $\pisces$
(Pisces).}  In many cases Kepler determines also the latitude, but
these data appear not to be very relevant for the discussion
concerning secular terms, so I omit them.

\figure{osserv.2}{\psfig{figure=kep_sec_gs.tps,width=15truecm}}{Difference
between the longitudes of Jupiter and Saturn calculated by Kepler from
the {\corsivo Tabul{\ae} Rudolphin{\ae}} and the positions evaluated
from the observations of Regiomontanus and Walther, vs.~the
observation date.  The plotted values come from the last column of
tables~\tabref{oss.saturno} and~\tabref{oss.giove}.  The graph clearly
shows that Jupiter seems to move faster than predicted, while Saturn
is slower.}

\table{oss.marte}{Observed positions of Mars compared with
the calculated ones.}{
\def\num#1{{\oldstyle #1}}
\vbox{\tabskip=0pt\offinterlineskip
\parindent 0pt\tabskip 0pt\halign{
\vrule$\altriga\ $\hfil{#}\ \vrule
&$\ $\hfil$\displaystyle{#}$\hfil{\ }
&{#}\hfil\ \vrule
&$\ $\hfil$\displaystyle{#}$\hfil{\ }
&$\ ${#}\hfil\ \vrule
&$\ $\hfil$\displaystyle{#}$\hfil\ \vrule
\cr
\noalign{\hrule}
Date\hfil  & {\rm Observed}  && {\rm Calculated} && {\rm Difference} \cr
\noalign{\hrule}
 2.12.1461 & {\num{27}}^{\circ}\,{\num{5}}'{\scriptstyle\frac{1}{2}}   & {\capricornus}
           & {\num{28}}^{\circ}\,{\num{4}}'                            & {\capricornus}
           & +\num{58}'{\scriptstyle\frac{1}{2}}
\cr				
24.12.1461 & {\num{14}}^{\circ}\,{\num{17}}'                           & {\aquarius}
           & {\num{15}}^{\circ}\,{\num{23}}'                           & {\aquarius}
           & +\num{1}^{\circ}\num{6}'
\cr				
15.09.1462 & {\num{21}}^{\circ}\,{\num{57}}'{\scriptstyle\frac{1}{2}}  & {\leo}
           & {\num{22}}^{\circ}\,{\num{1}}'                            & {\leo}
           & +\num{3}'{\scriptstyle\frac{1}{2}}
\cr				
 4.10.1462 & {\num{10}}^{\circ}\,{\num{56}}'                           & {\virgo}
           & {\num{10}}^{\circ}\,{\num{39}}'                           & {\virgo}
           & -\num{17}'
\cr				
11.12.1464 & {\num{28}}^{\circ}\,{\num{42}}'{\scriptstyle\frac{1}{2}}  & {\libra}
           & {\num{28}}^{\circ}\,{\num{29}}'                           & {\libra}
           & -\num{13}'{\scriptstyle\frac{1}{2}}
\cr				
19.06.1465 & {\num{1}}^{\circ}\,{\num{57}}'{\scriptstyle\frac{1}{2}}   & {\capricornus}
           & {\num{2}}^{\circ}\,{\num{13}}'                            & {\capricornus}
           & +\num{15}'{\scriptstyle\frac{1}{2}}
\cr				
26.04.1468 & {\num{29}}^{\circ}\,{\num{24}}'{\scriptstyle\frac{1}{2}}  & {\gemini}
           & {\num{29}}^{\circ}\,{\num{24}}'{\scriptstyle\frac{1}{2}}  & {\gemini}
           & \phantom{-}\num{0}'
\cr				
26.07.1471 & {\num{0}}^{\circ}\,{\num{22}}'                            & {\gemini}
           & {\num{0}}^{\circ}\,{\num{31}}'{\scriptstyle\frac{1}{2}}   & {\gemini}
           & +\num{9}'{\scriptstyle\frac{1}{2}}
\cr				
 7.03.1474 & {\num{0}}^{\circ}\,{\num{42}}'                            & {\leo}
           & {\num{1}}^{\circ}\,{\num{19}}'                            & {\leo}
           & +\num{37}'
\cr				
18.09.1475 & {\num{5}}^{\circ}\,{\num{54}}'                            & {\leo}
           & {\num{5}}^{\circ}\,{\num{38}}'{\scriptstyle\frac{1}{2}}   & {\leo}
           & -\num{15}'{\scriptstyle\frac{1}{2}}
\cr				
24.01.1476 & {\num{21}}^{\circ}\,{\num{44}}'                           & {\virgo}
           & {\num{22}}^{\circ}\,{\num{36}}'                           & {\virgo}
           & -\num{51}'{\scriptstyle\frac{1}{2}}
\cr
19.02.1476 & {\num{16}}^{\circ}\,{\num{53}}'                           & {\virgo}
           & {\num{16}}^{\circ}\,{\num{41}}'                           & {\virgo}
           & -\num{18}'
\cr				
12.04.1476 & {\num{5}}^{\circ}\,{\num{7}}'{\scriptstyle\frac{1}{3}}    & {\virgo}
           & {\num{4}}^{\circ}\,{\num{55}}'{\scriptstyle\frac{1}{2}}   & {\virgo}
           & -\num{11}'{\scriptstyle\frac{5}{6}}
\cr				
15.10.1477 & {\num{5}}^{\circ}\,{\num{49}}'                            & {\virgo}
           & {\num{5}}^{\circ}\,{\num{24}}'{\scriptstyle\frac{1}{3}}   & {\virgo}
           & -\num{24}'{\scriptstyle\frac{2}{3}}
\cr				
16.03.1478 & {\num{27}}^{\circ}\,{\num{6}}'{\scriptstyle\frac{1}{2}}   & {\libra}
           & {\num{27}}^{\circ}\,{\num{23}}'                           & {\libra}
           & +\num{16}'{\scriptstyle\frac{1}{2}}
\cr				
19.05.1478 & {\num{12}}^{\circ}\,{\num{7}}'{\scriptstyle\frac{1}{2}}   & {\libra}
           & {\num{12}}^{\circ}\,{\num{15}}'                           & {\libra}
           & -\num{3}'{\scriptstyle\frac{1}{2}}
\cr				
\noalign{\hrule}
\multispan 6{\altriga\hfil to be continued\qquad}\cr
}}}

\tabcont{oss.marte}{
\def\num#1{{\oldstyle #1}}
\vbox{\tabskip=0pt\offinterlineskip
\parindent 0pt\tabskip 0pt\halign{
\vrule$\altriga\ $\hfil{#}\ \vrule
&$\ $\hfil$\displaystyle{#}$\hfil{\ }
&{#}\hfil\ \vrule
&$\ $\hfil$\displaystyle{#}$\hfil{\ }
&$\ ${#}\hfil\ \vrule
&$\ $\hfil$\displaystyle{#}$\hfil\ \vrule
\cr
\noalign{\hrule}
Date\hfil  & {\rm Observed}  && {\rm Calculated} && {\rm Difference} \cr
\noalign{\hrule}
30.10.1479 & {\num{27}}^{\circ}\,{\num{24}}'                           & {\virgo}
           & {\num{27}}^{\circ}\,{\num{28}}'{\scriptstyle\frac{1}{2}}  & {\virgo}
           & +\num{4}'{\scriptstyle\frac{1}{2}}
\cr				
15.11.1479 & {\num{20}}^{\circ}\,{\num{56}}'                           & {\libra}
           & {\num{21}}^{\circ}\,{\num{7}}'{\scriptstyle\frac{1}{2}}   & {\libra}
           & +\num{11}'{\scriptstyle\frac{1}{2}}
\cr				
28.11.1484 & {\num{25}}^{\circ}\,{\num{52}}'                           & {\aries}
           & {\num{25}}^{\circ}\,{\num{34}}'{\scriptstyle\frac{1}{2}}  & {\aries}
           & -\num{18}'{\scriptstyle\frac{1}{2}}
\cr				
20.08.1486 & {\num{8}}^{\circ}\,{\num{0}}'                             & {\aries}
           & {\num{8}}^{\circ}\,{\num{4}}'                             & {\aries}
           & +\num{4}'
\cr				
 7.09.1486 & {\num{17}}^{\circ}\,{\num{8}}'                            & {\gemini}
           & {\num{17}}^{\circ}\,{\num{14}}'{\scriptstyle\frac{2}{3}}  & {\gemini}
           & +\num{6}'{\scriptstyle\frac{2}{3}}
\cr				
13.09.1488 & {\num{11}}^{\circ}\,{\num{21}}'                           & {\cancer}
           & {\num{11}}^{\circ}\,{\num{21}}'{\scriptstyle\frac{1}{2}}  & {\cancer}
           & +\num{0}'{\scriptstyle\frac{1}{2}}
\cr				
 3.10.1488 & {\num{21}}^{\circ}\,{\num{12}}'{\scriptstyle\frac{1}{2}}  & {\cancer}
           & {\num{21}}^{\circ}\,{\num{39}}'{\scriptstyle\frac{1}{3}}  & {\cancer}
           & +\num{26}'{\scriptstyle\frac{5}{6}}
\cr				
18.09.1490 & {\num{0}}^{\circ}\,{\num{12}}'{\scriptstyle\frac{1}{2}}   & {\cancer}
           & {\num{0}}^{\circ}\,{\num{15}}'{\scriptstyle\frac{1}{2}}   & {\cancer}
           & +\num{3}'
\cr				
23.03.1491 & {\num{22}}^{\circ}\,{\num{37}}'                           & {\leo}
           & {\num{22}}^{\circ}\,{\num{31}}'                           & {\leo}
           & -\num{6}'
\cr				
26.09.1492 & {\num{19}}^{\circ}\,{\num{9}}'{\scriptstyle\frac{1}{2}}   & {\leo}
           & {\num{19}}^{\circ}\,{\num{13}}'{\scriptstyle\frac{1}{2}}  & {\leo}
           & +\num{6}'
\cr				
27.03.1493 & {\num{3}}^{\circ}\,{\num{2}}'{\scriptstyle\frac{1}{2}}    & {\libra}
           & {\num{3}}^{\circ}\,{\num{11}}'                            & {\libra}
           & +\num{8}'{\scriptstyle\frac{1}{2}}
\cr				
30.09.1503 & {\num{11}}^{\circ}\,{\num{33}}'{\scriptstyle\frac{1}{3}}  & {\cancer}
           & {\num{11}}^{\circ}\,{\num{57}}'                           & {\cancer}
           & +\num{23}'{\scriptstyle\frac{2}{3}}
\cr				
 6.10.1503 & {\num{14}}^{\circ}\,{\num{23}}'{\scriptstyle\frac{1}{2}}  & {\cancer}
           & {\num{14}}^{\circ}\,{\num{34}}'{\scriptstyle\frac{1}{2}}  & {\cancer}
           & +\num{11}'{\scriptstyle\frac{1}{2}}
\cr				
16.11.1503 & {\num{23}}^{\circ}\,{\num{30}}'                           & {\cancer}
           & {\num{23}}^{\circ}\,{\num{59}}'                           & {\cancer}
           & +\num{29}'
\cr
19.01.1504 & {\num{6}}^{\circ}\,{\num{31}}'{\scriptstyle\frac{1}{2}}   & {\cancer}
           & {\num{6}}^{\circ}\,{\num{17}}'{\scriptstyle\frac{2}{3}}   & {\cancer}
           & -\num{13}'{\scriptstyle\frac{5}{6}}
\cr				
 6.02.1504 & {\num{4}}^{\circ}\,{\num{57}}'{\scriptstyle\frac{1}{2}}   & {\cancer}
           & {\num{4}}^{\circ}\,{\num{56}}'                            & {\cancer}
           & -\num{1}'{\scriptstyle\frac{1}{2}}
\cr				
 3.03.1504 & {\num{10}}^{\circ}\,{\num{57}}'                           & {\cancer}
           & {\num{11}}^{\circ}\,{\num{10}}'                           & {\cancer}
           & +\num{13}'
\cr				
30.05.1504 & {\num{20}}^{\circ}\,{\num{4}}' {\scriptstyle\frac{1}{2}}  & {\leo}
           & {\num{20}}^{\circ}\,{\num{11}}'{\scriptstyle\frac{1}{2}}  & {\leo}
           & +\num{7}'{\scriptstyle\frac{1}{2}}
\cr
\noalign{\hrule}
}}}

\table{oss.venere}{Observed positions of Venus compared with
the calculated ones.}{
\def\num#1{\obeyspaces{\oldstyle #1}}
\vbox{\tabskip=0pt\offinterlineskip
\parindent 0pt\tabskip 0pt\halign{
\vrule$\altriga\ $\hfil{#}\ \vrule
&$\ $\hfil$\displaystyle{#}$\hfil{\ }
&{#}\hfil\ \vrule
&$\ $\hfil$\displaystyle{#}$\hfil{\ }
&$\ ${#}\hfil\ \vrule
&$\ $\hfil$\displaystyle{#}$\hfil\ \vrule
\cr
\noalign{\hrule}
Date\hfil  & {\rm Observed}  && {\rm Calculated} && {\rm Difference} \cr
\noalign{\hrule}
14.12.1461 & {\num{ 0}}^{\circ}\,{\num{ 7}}'                           & {\aquarius}
           & {\num{ 0}}^{\circ}\,{\num{44}}'                           & {\aquarius}
           & +\num{37}'
\cr				
idem       & {\num{ 0}}^{\circ}\,{\num{31}}'                           & {\aquarius}
           & {\num{ 0}}^{\circ}\,{\num{44}}'                           & {\aquarius}
           & +\num{13}'
\cr
10.01.1462 & {\num{ 4}}^{\circ}\,{\num{52}}'                           & {\pisces}
           & {\num{ 4}}^{\circ}\,{\num{ 3}}'                           & {\pisces}
           & -\num{49}'		
\cr
idem       & {\num{ 4}}^{\circ}\,{\num{ 5}}'                           & {\pisces}
           & {\num{ 4}}^{\circ}\,{\num{ 3}}'                           & {\pisces}
           & -\num{ 2}'
\cr
idem       & {\num{ 3}}^{\circ}\,{\num{22}}'                           & {\pisces}
           & {\num{ 4}}^{\circ}\,{\num{ 3}}'                           & {\pisces}
           & +\num{41}'
\cr
19.09.1462 & {\num{20}}^{\circ}\,{\num{16}}'                           & {\leo}
           & {\num{20}}^{\circ}\,{\num{44}}'                           & {\leo}
           & +\num{28}'		
\cr
26.09.1462 & {\num{28}}^{\circ}\,{\num{42}}'                           & {\leo}
           & {\num{28}}^{\circ}\,{\num{35}}'                           & {\leo}
           & -\num{ 6}'		
\cr
20.10.1462 & {\num{27}}^{\circ}\,{\num{37}}'                           & {\virgo}
           & {\num{26}}^{\circ}\,{\num{45}}'{\scriptstyle\frac{1}{2}}  & {\virgo}
           & +\num{ 8}'{\scriptstyle\frac{1}{2}}
\cr				
25.10.1462 & {\num{ 2}}^{\circ}\,{\num{38}}'                           & {\libra}
           & {\num{ 2}}^{\circ}\,{\num{47}}'                           & {\libra}
           & +\num{ 9}'
\cr				
21.02.1478 & {\num{24}}^{\circ}\,{\num{50}}'                           & {\aries}
           & {\num{25}}^{\circ}\,{\num{ 0}}'                           & {\aries}
           & +\num{10}'
\cr				
11.08.1478 & {\num{10}}^{\circ}\,{\num{51}}'{\scriptstyle\frac{1}{2}}  & {\cancer}
           & {\num{10}}^{\circ}\,{\num{48}}'                           & {\cancer}
           & -\num{ 3}'{\scriptstyle\frac{1}{2}}
\cr				
15.11.1481 & {\num{16}}^{\circ}\,{\num{54}}'                           & {\libra}
           & {\num{16}}^{\circ}\,{\num{29}}'                           & {\libra}
           & -\num{25}'
\cr				
idem       & {\num{16}}^{\circ}\,{\num{46}}'                           & {\libra}
           & {\num{16}}^{\circ}\,{\num{29}}'                           & {\libra}
           & -\num{17}'
\cr
19.11.1481 & {\num{21}}^{\circ}\,{\num{22}}'                           & {\libra}
           & {\num{20}}^{\circ}\,{\num{57}}'{\scriptstyle\frac{1}{2}}  & {\libra}
           & -\num{24}'{\scriptstyle\frac{1}{2}}
\cr				
25.11.1481 & {\num{27}}^{\circ}\,{\num{52}}'                           & {\libra}
           & {\num{28}}^{\circ}\,{\num{ 0}}'                           & {\libra}
           & -\num{ 3}'
\cr				
idem       & {\num{28}}^{\circ}\,{\num{14}}'                           & {\libra}
           & {\num{28}}^{\circ}\,{\num{ 0}}'                           & {\libra}
           & +\num{14}'
\cr
20.09.1486 & {\num{23}}^{\circ}\,{\num{ 6}}'                           & {\leo}
           & {\num{23}}^{\circ}\,{\num{ 4}}'                           & {\leo}
           & -\num{ 2}'
\cr				
24.09.1486 & {\num{27}}^{\circ}\,{\num{30}}'{\scriptstyle\frac{3}{4}}  & {\leo}
           & {\num{27}}^{\circ}\,{\num{54}}'                           & {\leo}
           & +\num{24}'{\scriptstyle\frac{3}{4}}
\cr				
\noalign{\hrule}
\multispan 6{\altriga\hfil to be continued\qquad}\cr
}}}

\tabcont{oss.venere}{
\def\num#1{{\oldstyle #1}}
\vbox{\tabskip=0pt\offinterlineskip
\parindent 0pt\tabskip 0pt\halign{
\vrule$\altriga\ $\hfil{#}\ \vrule
&$\ $\hfil$\displaystyle{#}$\hfil{\ }
&{#}\hfil\ \vrule
&$\ $\hfil$\displaystyle{#}$\hfil{\ }
&$\ ${#}\hfil\ \vrule
&$\ $\hfil$\displaystyle{#}$\hfil\ \vrule
\cr
\noalign{\hrule}
Date\hfil  & {\rm Observed}  && {\rm Calculated} && {\rm Difference} \cr
\noalign{\hrule}
 1.04.1489 & {\num{27}}^{\circ}\,{\num{54}}'                           & {\taurus}
           & {\num{28}}^{\circ}\,{\num{ 6}}'                           & {\taurus}
           & +\num{12}'
\cr				
13.12.1490 & {\num{15}}^{\circ}\,{\num{45}}'                           & {\aquarius}
           & {\num{15}}^{\circ}\,{\num{50}}'                           & {\aquarius}
           & +\num{ 5}'
\cr				
17.01.1491 & {\num{23}}^{\circ}\,{\num{26}}'                           & {\aquarius}
           & {\num{23}}^{\circ}\,{\num{35}}'{\scriptstyle\frac{1}{2}}  & {\aquarius}
           & +\num{ 9}'{\scriptstyle\frac{1}{2}}
\cr				
14.02.1491 & {\num{15}}^{\circ}\,{\num{34}}'                           & {\aries}
           & {\num{16}}^{\circ}\,{\num{10}}'                           & {\aries}
           & +\num{36}'
\cr				
19.09.1494 & {\num{22}}^{\circ}\,{\num{30}}'                           & {\leo}
           & {\num{22}}^{\circ}\,{\num{27}}'{\scriptstyle\frac{1}{2}}  & {\leo}
           & -\num{ 2}'{\scriptstyle\frac{1}{2}}
\cr				
10.12.1503 & {\num{ 2}}^{\circ}\,{\num{ 4}}'{\scriptstyle\frac{1}{2}}  & {\aquarius}
           & {\num{ 2}}^{\circ}\,{\num{ 3}}'                           & {\aquarius}
           & -\num{ 1}'{\scriptstyle\frac{1}{2}}
\cr				
19.01.1504 & {\num{17}}^{\circ}\,{\num{36}}'{\scriptstyle\frac{1}{2}}  & {\capricornus}
           & {\num{18}}^{\circ}\,{\num{12}}'                           & {\capricornus}
           & +\num{35}'{\scriptstyle\frac{1}{2}}
\cr
24.01.1504 & {\num{16}}^{\circ}\,{\num{51}}'{\scriptstyle\frac{1}{2}}  & {\capricornus}
           & {\num{17}}^{\circ}\,{\num{ 6}}'                           & {\capricornus}
           & +\num{14}'{\scriptstyle\frac{1}{2}}
\cr				
27.01.1504 & {\num{16}}^{\circ}\,{\num{41}}'{\scriptstyle\frac{1}{2}}  & {\capricornus}
           & {\num{17}}^{\circ}\,{\num{ 4}}'                           & {\capricornus}
           & +\num{22}'{\scriptstyle\frac{1}{2}}
\cr				
20.02.1504 & {\num{26}}^{\circ}\,{\num{40}}'                           & {\capricornus}
           & {\num{27}}^{\circ}\,{\num{ 1}}'                           & {\capricornus}
           & +\num{21}'
\cr				
 3.03.1504 & {\num{ 6}}^{\circ}\,{\num{26}}'                           & {\aquarius}
           & {\num{ 6}}^{\circ}\,{\num{39}}'                           & {\aquarius}
           & +\num{13}'
\cr				
11.03.1504 & {\num{13}}^{\circ}\,{\num{58}}'                           & {\aquarius}
           & {\num{13}}^{\circ}\,{\num{54}}'                           & {\aquarius}
           & -\num{ 4}'
\cr				
12.03.1504 & {\num{15}}^{\circ}\,{\num{ 0}}'                           & {\aquarius}
           & {\num{14}}^{\circ}\,{\num{50}}'                           & {\aquarius}
           & -\num{10}'
\cr				
17.03.1504 & {\num{19}}^{\circ}\,{\num{46}}'{\scriptstyle\frac{1}{2}}  & {\aquarius}
           & {\num{19}}^{\circ}\,{\num{41}}'{\scriptstyle\frac{1}{2}}  & {\aquarius}
           & -\num{ 5}'
\cr				
18.03.1504 & {\num{20}}^{\circ}\,{\num{32}}'                           & {\aquarius}
           & {\num{20}}^{\circ}\,{\num{40}}'{\scriptstyle\frac{1}{2}}  & {\aquarius}
           & +\num{7}'{\scriptstyle\frac{1}{2}}
\cr				
19.03.1504 & {\num{21}}^{\circ}\,{\num{23}}'                           & {\aquarius}
           & {\num{21}}^{\circ}\,{\num{39}}'{\scriptstyle\frac{2}{3}}  & {\aquarius}
           & +\num{13}'{\scriptstyle\frac{2}{3}}
\cr				
27.03.1504 & {\num{29}}^{\circ}\,{\num{50}}'                           & {\aquarius}
           & {\num{29}}^{\circ}\,{\num{51}}'                           & {\aquarius}
           & +\num{1}'
\cr
\noalign{\hrule}
}}}

\table{oss.mercurio}{Observed positions of Mercury compared with
the calculated ones.}{
\def\num#1{\obeyspaces{\oldstyle #1}}
\vbox{\tabskip=0pt\offinterlineskip
\parindent 0pt\tabskip 0pt\halign{
\vrule$\altriga\ $\hfil{#}\ \vrule
&$\ $\hfil$\displaystyle{#}$\hfil{\ }
&{#}\hfil\ \vrule
&$\ $\hfil$\displaystyle{#}$\hfil{\ }
&$\ ${#}\hfil\ \vrule
&$\ $\hfil$\displaystyle{#}$\hfil\ \vrule
\cr
\noalign{\hrule}
Date\hfil  & {\rm Observed}  && {\rm Calculated} && {\rm Difference} \cr
\noalign{\hrule}
22.10.1481 & {\num{19}}^{\circ}\,{\num{12}}'                           & {\libra}
           & {\num{19}}^{\circ}\,{\num{ 3}}'                           & {\libra}
           & -\num{ 9}'
\cr				
idem       & {\num{19}}^{\circ}\,{\num{22}}'                           & {\libra}
           & {\num{19}}^{\circ}\,{\num{ 3}}'                           & {\libra}
           & -\num{19}'
\cr
 3.11.1481 & {\num{ 4}}^{\circ}\,{\num{36}}'                           & {\scorpio}
           & {\num{ 5}}^{\circ}\,{\num{ 2}}'                           & {\scorpio}
           & +\num{26}'
\cr				
idem       & {\num{ 4}}^{\circ}\,{\num{46}}'                           & {\scorpio}
           & {\num{ 5}}^{\circ}\,{\num{ 2}}'                           & {\scorpio}
           & +\num{36}'
\cr
11.10.1482 & {\num{ 9}}^{\circ}\,{\num{32}}'{\scriptstyle\frac{1}{2}}  & {\libra}
           & {\num{ 9}}^{\circ}\,{\num{27}}'                           & {\libra}
           & -\num{ 5}'{\scriptstyle\frac{1}{2}}
\cr				
16.01.1488 & {\num{23}}^{\circ}\,{\num{40}}'                           & {\aquarius}
           & {\num{23}}^{\circ}\,{\num{40}}'                           & {\aquarius}
           & \num{ 0}'
\cr				
26.08.1491 & {\num{23}}^{\circ}\,{\num{14}}'                           & {\leo}
           & {\num{23}}^{\circ}\,{\num{13}}'                           & {\leo}
           & -\num{ 1}'
\cr				
30.08.1491 & {\num{27}}^{\circ}\,{\num{10}}'                           & {\leo}
           & {\num{27}}^{\circ}\,{\num{19}}'                           & {\leo}
           & +\num{ 9}'
\cr				
31.08.1491 & {\num{28}}^{\circ}\,{\num{34}}'                           & {\leo}
           & {\num{28}}^{\circ}\,{\num{38}}'                           & {\leo}
           & +\num{ 4}'
\cr				
 2.09.1491 & {\num{ 1}}^{\circ}\,{\num{17}}'                           & {\virgo}
           & {\num{ 1}}^{\circ}\,{\num{30}}'{\scriptstyle\frac{1}{2}}  & {\virgo}
           & +\num{13}'{\scriptstyle\frac{1}{2}}
\cr				
 3.09.1491 & {\num{ 3}}^{\circ}\,{\num{ 9}}'                           & {\virgo}
           & {\num{ 3}}^{\circ}\,{\num{ 3}}'                           & {\virgo}
           & -\num{ 6}'
\cr				
 9.09.1491 & {\num{13}}^{\circ}\,{\num{27}}'                           & {\virgo}
           & {\num{13}}^{\circ}\,{\num{21}}'                           & {\virgo}
           & -\num{ 6}'
\cr				
 9.01.1504 & {\num{ 3}}^{\circ}\,{\num{36}}'{\scriptstyle\frac{1}{2}}  & {\capricornus}
           & {\num{ 3}}^{\circ}\,{\num{38}}'                           & {\capricornus}
           & +\num{ 1}'{\scriptstyle\frac{1}{2}}
\cr				
10.01.1504 & {\num{ 4}}^{\circ}\,{\num{21}}'{\scriptstyle\frac{1}{2}}  & {\capricornus}
           & {\num{ 4}}^{\circ}\,{\num{47}}'                           & {\capricornus}
           & +\num{25}'{\scriptstyle\frac{1}{2}}
\cr				
11.03.1504 & {\num{17}}^{\circ}\,{\num{34}}'{\scriptstyle\frac{1}{2}}  & {\aries}
           & {\num{17}}^{\circ}\,{\num{39}}'                           & {\aries}
           & +\num{ 4}'{\scriptstyle\frac{1}{2}}
\cr				
17.03.1504 & {\num{25}}^{\circ}\,{\num{34}}'{\scriptstyle\frac{1}{2}}  & {\aries}
           & {\num{25}}^{\circ}\,{\num{36}}'                           & {\aries}
           & +\num{ 1}'{\scriptstyle\frac{1}{2}}
\cr				
18.03.1504 & {\num{26}}^{\circ}\,{\num{24}}'{\scriptstyle\frac{1}{2}}  & {\aries}
           & {\num{26}}^{\circ}\,{\num{36}}'                           & {\aries}
           & +\num{ 1}'{\scriptstyle\frac{1}{2}}
\cr				
24.03.1504 & {\num{ 0}}^{\circ}\,{\num{ 9}}'{\scriptstyle\frac{1}{2}}  & {\taurus}
           & {\num{ 0}}^{\circ}\,{\num{29}}'                           & {\taurus}
           & +\num{19}'
\cr
idem       & {\num{ 0}}^{\circ}\,{\num{29}}'{\scriptstyle\frac{1}{2}}  & {\taurus}
           & {\num{ 0}}^{\circ}\,{\num{29}}'                           & {\taurus}
           & -\num{ 0}'{\scriptstyle\frac{1}{2}}
\cr
\noalign{\hrule}
}}}

\figure{osserv.3}{\psfig{figure=kep_sec_mvm.tps,width=15.5truecm}}{Difference
between the longitudes of Mercury, Venus and Mars calculated by Kepler from
the {\corsivo Tabul{\ae} Rudolphin{\ae}} and the positions evaluated
from the observations of Regiomontanus and Walther. In contrast with
fig.~\figref{osserv.2} no systematic deviation is observed here.}

A plot of the last column in the tables, namely the differences
between the calculated positions and the observed ones, is reported in
fig.~\figref{osserv.2} for Jupiter and Saturn and in
fig.~\figref{osserv.3} for Mercury, Venus and Mars.  The observational
errors are quite big, which is not surprising if one recalls the few
examples that I have given; thus the data point have a significant
dispersion.  However, the first figure clearly exhibits a systematic
deviation for Jupiter ad Saturn: the difference for Jupiter is always
positive (about $+18'$ according to Kepler), while for Saturn it is
always negative (about $-43'$).  That is, Jupiter moves faster
than predicted, while Saturn is slower.  No systematic deviation 
occurs instead for the three smaller planets, although Kepler's
claims that some exists also for Mars.

\section{kepsec.2}{The need for secular equations}
From Kepler's considerations it seems that he concluded for the
necessity of secular equations\footnote{It should be stressed here
that the meaning assigned by Kepler to the word ``equation''
corresponds rather to what we call ``function''. In fact he was
looking for a (periodic) correction of the mean motions with respect
to a fixed mean value.} after a very careful examination of the last
observation of a great conjunction between Jupiter and Saturn, by
Walther.  Walther's notes report that he observed the conjunction in
the evening of may 24, 1504 (he died on June 19).  However, according
to Kepler's calculations the distance between the two planets should
have been about $1^{\circ}$ at that time.  Kepler writes:\footnote{And
this discrepancy in the calculation for Jupiter and Saturn, which
amounts to a whole degree, is such an obstacle that it caused me to be
assailed by many perplexities, and for five solid months I have been
troubled until I eventually came to think about average motions in a
different way, having accepted the manifest secular inequality of the
motions.  (I came to this conclusion on June 18, 1624.)''}

\citazione
Et hic dissensus calculi in $\jupiter$ and $\saturn\,$, excurrens ad
integrum gradum, est remora illa, qu{\ae} me, plurima perplexitate
circumventum, per solidos quinque menses in observationibus
Waltherianis exercuit tandemque ad nova consilia circa motuum mediorum
speculationem adegit, deprehensa manifesta inequalitate motuum
seculari.  (Absolvi hucusque 18 Junii 1624.) 
\finecitazione

\noindent
Having thus decided that a secular equation is needed Kepler turns to
determining it.  Here he faces big difficulties, partly due to the
lack of data, partly to an hypothesis that he strongly wants to be
satisfied (although this may even appear as foolish to us).  Let me
discuss this part in some detail: this will enlight which secular
equation he was looking for.

Kepler starts by examining five ancient data for Saturn reported by
Ptolemy.  The first one refers to an observation made by Chaldean
astronomers:\footnote{``$\ldots\>$in the evening of the second day of
the Xantic month of the year 82, which according to the interpreters
of Ptolemy is March 1, 229 BC.  Then Saturn was seen two fingers below
the austral shoulder of Virgo''.}  {\corsivo ``$\ldots\>$a. 82 die 2
Xantichi vesperi, quod ex fide Ptolem{\ae}i interpretis fuit ante
Chr. anno 229 d. 1 Mart.  Tunc {\saturn} sub australi humero {\virgo}
visus est 2 digitos''}.  Four more observations are due to Ptolemy
himself.  Three of them are {\grm\char'202 kronuqoi} (i.e., with no
time indication) and give the calculated oppositions of Saturn with
the average Sun that occurred in $1^{\circ}\,13'\>\libra$,
$9^{\circ}\,40'\>\sagittarius$ and $14^{\circ}\,14'\>\capricornus\,$.
The fourth one reports that that Saturn was observed\footnote{``at
$9^{\circ}\,15'\>\aquarius$, when the Moon was seen half a degree
forward, i.e., at $9^{\circ}\,45'\>\aquarius\,$; Saturn was
observed with respect to the star named `clara hyadum', but the
circumstances are fallacious.'' (The star 'clara hyadum' is
Aldebaran.)} {\corsivo ``in $9^{\circ}\,15'\>\aquarius$, quando
$\rightmoon$ est observata dimidio gradus ultra, i.e.~in
$9^{\circ}\,45'\>\aquarius\,$; observatus est $\saturn$ ad claram
Hyadum, sed lubric{\ae} sunt circumstanti{\ae}''.}

First Kepler recalculates the longitudes of the oppositions with the
true position of the Sun, also taking into account the renumbering of
zodiacal signs with respect to Ptolemy's time, which according to
Kepler corresponds to $1^{\circ}\,3'\,$.  Adding both corrections he
finds $2^{\circ}\,28'\>\libra$, $10^{\circ}\,43'\>\sagittarius$ and
$15^{\circ}\,10'\>\capricornus\,$.  Then he keeps fixed the
eccentricity of Saturn's orbit, {\corsivo ``qu{\ae} hodie ex
accuratissimis observationibus 30 continuorum annorum stabilitur''}
(which today is well established in view of 30 continuous years of
very accurate observations), and by trying different positions of the
aphelion he calculates the mean motion that gives him the best
approximation of Ptolemy's longitudes.  Here, he clearly assumes that
the Earth's motion is uniform over all centuries, and calculates the
intervals between two oppositions by Saturn's motion on the ellipse.
This represents a first partial success, since it allows him to fit
the observations within $10'$.  After that he comes to fit also the
fourth observation.  He can determine quite accurately the longitude
and the time thanks to the position of the Moon, which is reported by
Ptolemy.

\looseness -1
Finally, he tries to fit also the Chaldean observation, and he
succeeds again, but with a further minor change of the mean motion.
Here are Kepler's results for the five observations; the first line
gives the predictions, the second one the data recalculated by Kepler
by including the corrections mentioned above.  For the fourth
observation of Ptolemy he gives two possible corrections, the second
one based on Moon's position.
$$
\vcenter{\openup1\jot\halign{
\hfil$\displaystyle{#}$
&$\displaystyle{#}$\hfil\quad
&\hfil$\displaystyle{#}$
&$\displaystyle{#}$\hfil\quad
&\hfil$\displaystyle{#}$
&$\displaystyle{#}$\hfil\quad
&\hfil$\displaystyle{#}$
&$\displaystyle{#}$\hfil\quad
&\hfil$\displaystyle{#}$
&$\displaystyle{#}$\hfil\quad
&$\displaystyle{#}$\hfil
\cr
1^{\circ}\,36'\,&\libra\>, & 10^{\circ}\,38'\,&\sagittarius\>, &
 15^{\circ}\,10'\,&\capricornus\>, & 10^{\circ}\,22'\,&\aquarius\>, &
  8^{\circ}\,56'\,23''\,&\virgo\>,
& {\rm pro}
\cr
2^{\circ}\,16'\,&\libra\>, & 10^{\circ}\,43'\,&\sagittarius\>, &
 15^{\circ}\,17'\,&\capricornus\>, & 10^{\circ}\,18'\,&\aquarius\>, &
  8^{\circ}\,41'\,\phantom{0}0''\,&\virgo\>,
\cr
&&&&& {\rm vel} & 10^{\circ}\,27'\, & \multispan 1{${\rm ex}\rightmoon\>.$}
\cr
}}
$$
Kepler's goal in this calculation is to distribute the deviation, and
indeed the four observations of Ptolemy turn out to be all in defect,
while the Chaldean one is in excess.  On the other hand, he remarks
that taking out the Chaldean observation he can uniformly distribute
the deviations among the four observation of Ptolemy within $9'$, but
again he must change the mean motion and move the aphelion by about
$2^{\circ}$.  This represents a partial success that Kepler appears to
be very proud of, since he says: {\corsivo ``Conciliet eas propius,
qui id potest, salva commensuratione orbium, per Tychonicas
certissimas inventa''} (reconcile them better, who can, provided the
proportions among the orbits are respected, since we have firmly
found them from Tycho).

But here the major troubles show up.  Kepler realizes that he can find
an appropriate secular equation for a given, not too long time
interval, but he is unable to find one which is valid for all times.
Coming indeed to recent observations (for him) so he
writes.\footnote{``Moreover, using their epochs I have calculated the
observations of Walther and Regiomontanus, and from them undoubtedly
appears that the motion of Saturn is affected by a secular equation.
Thus it is vain that we try to find an average between very distant
observations, if they fight together and do not accept to be
represented by a definite proportion confirmed by reliable and close
observations; ($\ldots$) Concerning Tycho's time indeed it seems to me
that considering achronichous oppositions over a complete period of
thirty years the effect of a secular equation faintly appears.  This
however usually does not happen for a maximal equation, since in that
case the quantity remains almost constant due to an insensible change,
but it rather happens close to a null value, when the quantities to be
added before and to be subtracted after, or the contrary, reach the
maximum attainable.''}

\citazione
At cum ex his epochis computarem postea Waltherianas et Regiomontani
observationes exque iis appareret clarissime, $\saturn$ motus indigere
{\ae}quatione seculari, eoque frustra nos medium affectare inter longe
distantes, si inter se pugnent, nec in unam certis vicinis
observationibus confirmatam commensurationem se cogi patiantur;

$\ldots\quad\ldots$

\noindent
Nam quod Tychonicum attinet, videor ex oppositionibus acronychiis per
totam triaconta\"ederis periodum jam sentiscere effectum {\ae}quationis
secularis.  Id autem fieri solet non in {\ae}quatione maxima, tunc enim
quantitas consistit, insensibili existente varietate, sed in
{\ae}quatione prope nulla, tunc enim desinente adjectoria, incipiente
subtratoria, vel e contrario, quantum potest maxima sentitur.
\finecitazione

\noindent
Thus, the first conclusion made by Kepler is that {\corsivo the
secular equation can be calculated only relative to a short fixed
period}.

Let me add a remark.  Kepler seems to assume that the planets
possess a proper mean motion, and the secular equation consists in a
periodic oscillation of the actual mean motion around the proper value
that can be represented, e.g., by a sine function.  What Kepler is
attempting is to evaluate such a function (namely the mean velocity)
using experimental data, but the lack of data does not allow him to
complete his job. Moreover he clearly misses all technical tools that
have been developed more than half a century later, starting with the
{\corsivo Methodus fluxionum et serierum infinitorum} of Isaac Newton
(1643--1727).  His remark concerning Tycho's time is essentially
that the derivative of a function such as, e.g., the sine takes a
maximum where its value is zero, while the function is almost constant
in the vicinity of a maximum or minimum.

\figure{kepsec.1}{\psfig{figure=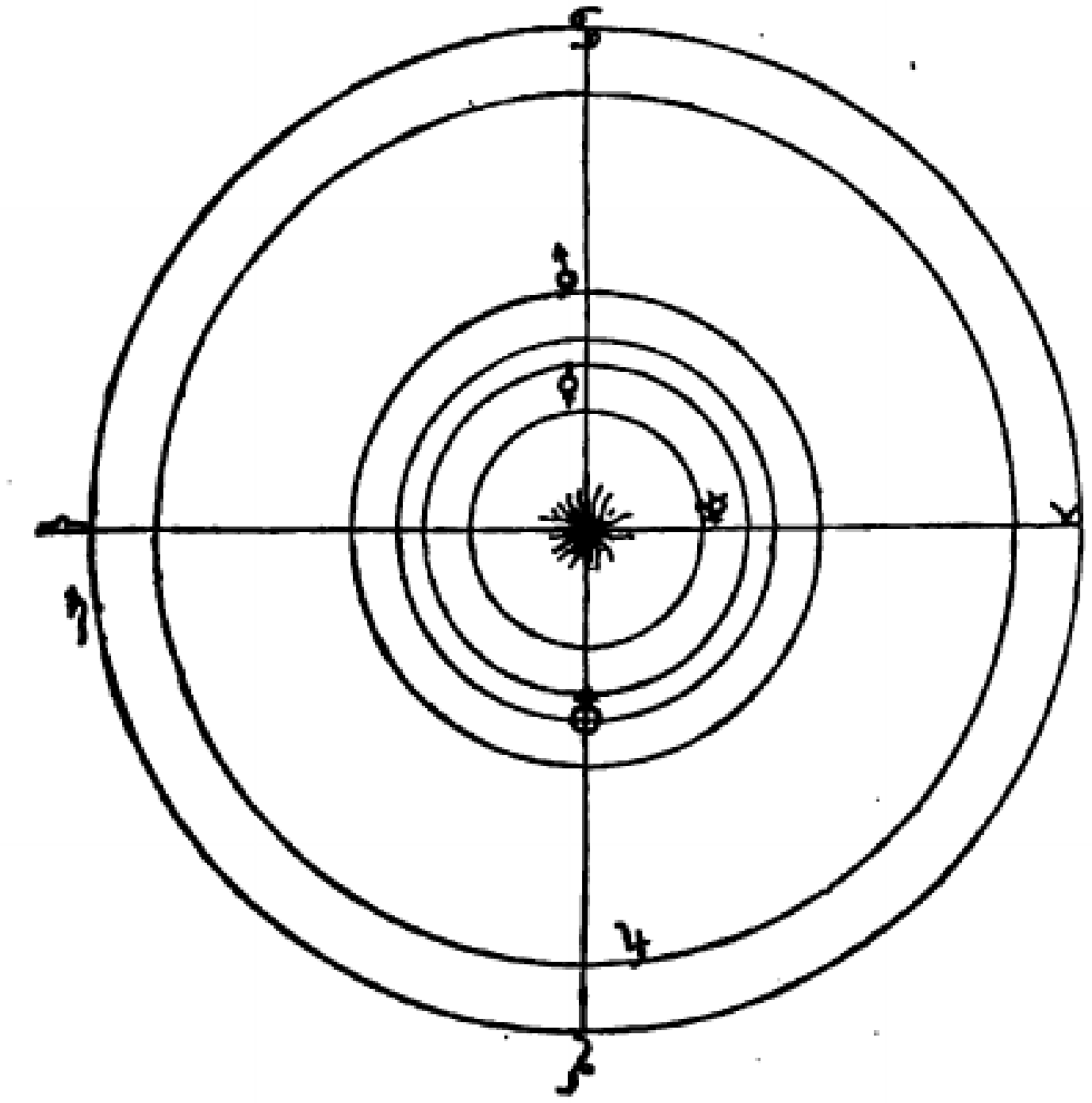,height=12truecm}}{The
configuration of the Solar System at creation time, according to one
of the attempts by Kepler to place the planets at the cardinal points
of the orbit of the Earth. The planets are close to the beginning of
the constellations of $\aries$ (Aries), $\cancer$ (Cancer), $\libra$
(Libra) and $\capricornus$ (Capricornus). (Figure taken from vol.~VI
of {\corsivo Johannis Kepleri Opera Omnia}, pag.~30.)}

Now the problem is: where from does the proper mean motion come, and
which is its value?  The consideration concerning the observations of
Tycho seem to suggest that Kepler's first idea was that the wanted value
could coincide with the observed one at Tycho's time (actually, he
later admits that this might be false).  However, here comes the extra
hypothesis that I have mentioned at the beginning of this section:
Kepler strongly believes that at the beginning of time (the creation)
the planets were in a privileged configuration, identified through the
cardinal points of the orbit of the Earth.  This is well stated in
ch.~XXIII of {\corsivo Mysterium Cosmographicum}:\footnote{It is
certain that God did not establish the motions inconsiderately, but
from one well definite beginning and a privileged configuration of
stars, and at the beginning of the zodiac, which has been moulded by
means of the inclination of the Earth, our house, because everything
has been created for the human beings.''}

\citazione
Certe non temere Deus instituit motus, sed ab uno quodam certo
principio et illustri stellarum conjunctione, et in initio zodiaci,
quod creator per inclinationem Telluris domicilii nostri effinxit,
quia omnia propter hominem.\finecitazione

\noindent
Kepler also attempted at determining the date of creation by combining
calculations based on Genesis (which was very common at that time,
typically giving dates around 4000~BC) and the configuration of the
planets as given by the tables.  In a note added to ch. XXIII of
{\corsivo Mysterium Cosmographicum} he claims that the Sun and the
Moon were created on July 24, 3993~BC.  Figure~\figref{kepsec.1} fairly
represents Kepler's thought.\footnote{Figure~\figref{kepsec.1} is
found in a letter of Kepler to Peter Cr\"uger (1580--1639) of February
18, 1624, where Kepler remarks that the calculated positions of the
planets are very close to the cardinal points, and that trying to
reduce them at the beginning of the zodiacal signs {\corsivo ``ego
nunc pene oculos ipsos computando perdo''} (I'm almost consuming my
eyes in computing).  As to the calculation of the creation date,
it may be curious to note that Jeremiah Horrocks (1618--1641) wrote in
1637 that compared to other similar calculations Kepler's one is
{\corsivo ``ingeniosa conjectura et non improbabilis''} (an ingenious
and non improbable conjecture).}  Besides the positions of the
planets, he also tried hardly to show that at the same times also the
aphelia and the nodes of the orbits were located at the same points,
in a beautiful geometrical arrangement.  This hypothesis actually
represents for Kepler a major obstacle in trying to determine the
secular equations.  As it appears from the figure, some planets are
not perfectly aligned with the cardinal points, what struggled Kepler
a lot.

Kepler's note contains a long discussion on the old observations of
Jupiter and Mars.  In particular he shows that a mere change in the
position of the aphelion will not explain the observed inequalities
for Jupiter.  Here he also raises several doubts concerning the
correspondence among different calendars.  Some of his doubts are
dictated by his difficulty in fitting not just the data, but the
initial configuration of the orbits.  I skip this part, since for the
purpose of justifying the need for secular equations it adds very
little to the considerations on Saturn that I have reported.  However,
it is interesting to quote a sort of conclusion which fully enlights
Kepler's claim, in the quotation at the beginning of this note, that
many centuries of observations will pass before we can discover which
secular equations we must introduce.  Having remarked that perhaps an
appropriate secular equation might fully remove the discrepancies with
the observations of Ptolemy and Dionysius he says:\footnote{``For he
who will succeed in doing it will also convey information to posterity
what must be added at Dionysius time, what at Ptolemy's time (since I
do exactly that for the equinoxes at Ptolemy's time), and will also
prescribe that they must annotate the excess or the defect for all
future centuries (that certainly will be observed for all eccentrics).
Then our descendants, on the basis of many data concerning excess and
defect, will eventually succeed in studying a circular and ordered
reconstruction and justifying it by calculations: and even, if needed,
to change the velocity of the motions, so as to appropriately place
our century in excess or in defect.''}

\citazione
Nam qui hoc fecerit, is admonitionem transmittere possit ad posteros,
quid tempore Dionysii addendum, quid tempore Ptolem{\ae}i (sicut hoc
ipsum ego facio in {\ae}quinotiis ad tempora Ptolem{\ae}i), jubens
etiam ceterorum sequentiom seculorum excessus vel defectus (qui quidem
per totos eccentricos conspiciantur) annotare, tandemque eos, qui
victuri sunt, ex pluribus idoneis defectibus et excessibus circularem
et ordinatam restitutionem, qualis procul dubio est, investigare
numerisque explicare: quin etiam, si aliter fieri nequeat, mutata
motus celeritate, ipsum etiam hoc nostrum seculum in excessu vel
defectu collocare.
\finecitazione

\noindent
It seems evident here that the ``secular equation'' sought by Kepler
is a periodic oscillation around an average value which reminds the
epicycles of classical astronomy.  His hope is that by accumulating
enough experimental points one will be able to calculate a periodic
function for the velocity.  Furthermore, his conclusion fully agrees
with the attitude of classical astronomers: {\corsivo the motion of
the planets can be explained in terms of superposition of periodic
motions, and the periods of such motions must be discovered through
observations}.  As to the average values of the motions, they should
coincide with the velocities at creation time.  But if we remove the
latter request and look only for the average observed values then we
get very close to the attitude of astronomers after Lagrange and
Laplace, as I will briefly illustrate in the next section.\footnote{It
may appear quite puzzling that Kepler does not explain how a varying
mean motion may be reconciled with his hypothesis (which he strongly
defends) that the proportions of the orbits and of the eccentricities
are fixed.  For, this seems to be in contrast with his third law,
according to which one has $n^2 a^3={\rm const}$, where $n$ is the
mean motion and $a$ the semimajor axis of the ellipse of a planet.
Although Kepler in his note does not mention this problem, it seems to
me that a straightforward interpretations is that the third law
applies to the {\corsivo initial} motion, which is assigned once for
all to every planet and coincides with the average mean motion.}

\section{kepsec.4}{After Kepler}
A thorough discussion of the development of our knowledge after Kepler
would not fit in a short note.  Here I give a brief sketch. For more
details see, e.g.,~\dbiref{Laskar-2006} or~\dbiref{Wilson-1985}.

The note of Kepler, although unpublished, did not remain completely
unknown: there are references in the literature to Kepler's
conclusions, but usually in a generic form, typically saying that
Kepler noticed remarkable inequalities in the motions of Saturn and
Jupiter, and that Jupiter appears to accelerate, while Saturn seems to
decelerate.  Actually, this is what Kepler himself had communicated
to his correspondents in some letters, without including further
details.  Furthermore, the significant increase of the number of
observations together with the improvement of the precision confirmed
more and more the existence of inequalities in the motion of the two
biggest planets, somehow fulfilling Kepler's desire that observational
data should be carefully collected.

The first attempt to include secular terms in tables is due to Edmond
Halley (1656--1742), who in 1719 published new tables including
secular corrections.  However, he did not exploit Kepler's conjecture
that one should look for a periodic behaviour (and likely he did not
know it, since Kepler's note was still unpublished at that time).
Pragmatically, he did just interpolate his data (around 1700) with
Ptolemy's, and introduced a linear increment for the mean motion of
Jupiter and a linear decrement for that of Saturn, that he called
{\corsivo secular}, as in Kepler.  This, he claimed, could be
enough in order to provide correct predictions for about 6000 years
before and after 1700.  Actually, a rough calculation based on his
data and taking into account the third Kepler's law leads to the
conclusion that about $2.7$ millions of years ago Jupiter ad Saturn
were on the same orbit, which is quite unlikely.  But probably at
Halley's time this was not a concern --- and definitely it is not if
one believes that the world was created less that 6000 years ago.
What remains true is that Halley's secular corrections (linear in time
for the mean motions) were taken as the paradigmatic reference for
half a century and more, thus completely ignoring Kepler's idea that
the secular motions should exhibit periods.

The accumulation of data showing with more and more evidence the
inequalities of the two biggest planets raised the question whether
the recent Newtonian theory of gravitation could be able to explain
these phenomena.  For it had been already pointed out by Newton that
the mutual attraction among planets could induce a slow change in the
orbits.  The French Academy proposed three prizes, in 1748, 1750 and
1752, for solving this question.  The 1748 and 1752 prizes were
awarded to Leonhard Euler (1707--1783), while the 1750 one was not
assigned.  Euler's memoirs had the merit of creating the skeleton of
perturbation theory, although, strictly speaking, his result was
wrong: he calculated secular terms, in Halley's sense (i.e., linear in
time for the mean motions), for the motion of Jupiter and Saturn
which, however, had the same sign for both
planets~\dbiref{Euler-1748}\dbiref{Euler-1752}.  Thus, according to
Euler's theory both planets should accelerate, in clear contrast with
the observations.  The error was partially corrected in 1762 by
Joseph--Louis Lagrange (1736--1813): at least, he got the right sign,
negative for Saturn and positive for Jupiter, still maintaining that
the secular term was a linear one on the mean
motion~\dbiref{Lagrange-1762}.  Eleven years later, in 1773, Pierre
Simon de Laplace (1749--1827) remarked that by improving the
approximation the secular terms found by Lagrange are canceled out by
other contributions that Lagrange had neglected~\dbiref{Laplace-1773}.
This question was eventually settled by Lagrange, who proved that no
secular variation (linear in time) of the semimajor axes can occur in
the approximation of first order in the masses of the planets, while a
periodic variation may occur.  Lagrange's result is remarkable, since
the secular invariance of the semimajor axes of the planetary orbits
represents a milestone in the research concerning the stability of the
Solar System.  Yet, his result cannot be considered as a conclusive
one, since it holds true only in the first order approximation on the
masses.  On the other hand, at that time it had an unpleasant
consequence: it implied that the ``secular terms'' introduced by
Halley are not justified by Newton's theory; this meant that the
problem of the inequalities of Jupiter and Saturn was reopened.  The
enigma was solved only in 1785, when Laplace discovered a long period
perturbation due to the approximate resonance $2\!:\!5$ between the
periods of Jupiter and Saturn~\dbiref{Laplace-1785}: this perturbation
has been named {\corsivo the great inequality}.  The period turns out
to be about 900 years.  Meanwhile Lagrange, soon followed by Laplace,
had also developed a theory for the secular motions of nodes and
perihelia together with the inclinations and the eccentricities of the
orbits~\dbiref{Lagrange-1774}\dbiref{Lagrange-1781}\dbiref{Lagrange-1782}\dbiref{Laplace-1773}.

The theory developed by Lagrange and Laplace may be considered in some
sense as a vindication of Kepler's intuition that the secular
perturbations should be periodic.  But it seems that nobody remarked
it.  On the other hand, two points deserve to be mentioned.  The first
one is that according to Lagrange and Laplace's theory {\corsivo all}
the orbital parameters are affected by periodic perturbations,
including the semimajor axes and the eccentricities that Kepler had
refused to change in view of the very precise observations of Tycho.
The second point is that we had not to wait for many centuries of
steady and careful observations, as predicted by Kepler.  Both these
points have a common explanation in the development of Newton's
gravitational theory: {\corsivo the frequencies of the perturbation
can be calculated on the basis of a mathematical model}.  Furthermore,
the theory offers a method for determining at one time the most
relevant deviations from the elliptic motions, because the equations
tie all these quantities together.  Indeed it is quite evident that
discovering the exact form of a perturbation with a period of 900
years on the basis of less than two centuries of observations is a
hard task, to say it fairly.  Furthermore, separating the different
contributions coming from slow changes in all parameters using only
his own intuition is clearly impossible, and this had been indeed
painful for Kepler.

\looseness -1
Thus Kepler's intuition that the secular inequalities should be
periodic seems to be confirmed by the theory.  But, once again,
reality is complex enough to escape our theories.  A considerable
amount of work has been spent during the XIX century in order to prove
that the theory of Lagrange remains true at any order in the masses:
the underlying idea is that the motion can be represented as a
superposition of periodic components --- the modern version of
epicycles, that we rename Fourier series or quasi--periodic motion.
On the other hand, the intuition that resonances could produce a
destroying effect on the regularity of the orbits began to appear:
this second aspect is connected with the hard problem of the so called
{\corsivo small divisors} (for an introductory reference see,
e.g.,~\dbiref{Giorgilli-1998}).

\looseness -1
The representation of the planetary dynamic as composed of
quasi--periodic motions has been seriously questioned at the end of
the XIX century.  On the one hand a brilliant Romanian student, Spiru
Haretu, proved that secular terms (in Halley's sense) do appear in
perturbation expansions at third order in the
masses~\dbiref{Haretu-1885}.  This result was soon forgotten, because
new expansion methods were developed which at least in some cases
could avoid such unwanted terms.  On the other hand, the celebrated
memoir of Poincar\'e on the problem of three bodies revealed that a
chaotic behaviour may show up.  This seems to be the end of the game,
but in 1954 Kolmogorov proved that quasi--periodic motions can
persist, and they represent the majority of the possible motions if
the masses of the planets are small enough~\dbiref{Kolmogorov-1954}.
But, what does ``small enough'' mean?  This remains an open
question~\dbiref{Gio-Loc-2009}.  It should also be mentioned that
massive long term calculations of the planetary orbits have revealed
that chaotic motions do indeed show up, although over time spans of
several millions of
years~\dbiref{Laskar-1989}\dbiref{Laskar-1994},\dbiref{Laskar-2009}.

\looseness -1
More than one century after Poincar\'e and more than fifty years after
Kolmogorov we are still unable to give a complete answer to the
problem of long term dynamics of the planets: the Solar System
seems to interweave ordered and chaotic behaviour in an inextricable
and intriguing manner.  But this is a long story that can not fit in
the present note.  I conclude by emphasizing that the problem raised
by Kepler is still there: we just succeeded in reformulating it in a
definitely more refined but more complicated way.

\references

\bye